\documentclass[useAMS]{mn2e}
\usepackage{epsfig}

\title[$z>5$ QSOs]{Expected $z>5$ QSO number counts in large area deep
  near-infrared surveys.}

\author[Fontanot, Somerville \& Jester]{Fabio Fontanot$^1$, Rachel S.
  Somerville$^1$ \& Sebastian Jester$^1$ \\
  $^1$ MPIA Max-Planck-Institute fuer Astronomie, Koenigstuhl 17, 69117 Heidelberg, Germany\\
  email: fontanot@mpia.de, somerville@mpia.de, jester@mpia.de}

\begin{document}
\date{Accepted ... Received ...}

\maketitle

\begin{abstract}
  The QSO luminosity function at $z>5$ provides strong constraints on
  models of joint evolution of QSO and their hosts. However, these
  observations are challenging because the low space densities of
  these objects necessitate surveying of large areas, in order to
  obtain statistically meaningful samples, while at the same time
  cosmological redshifting and dimming means that rather deep Near
  Infrared (NIR) imaging must be carried out. Several upcoming and
  proposed facilities with wide-field NIR imaging capabilities will
  open up this new region of parameter space. In this paper we present
  predictions for the expected number counts of $z>5$ QSOs, based on
  simple empirical and semi-empirical models of QSO evolution, as a
  function of redshift, depth and surveyed area.  We compute the
  evolution of observed-frame QSO magnitudes and colors in a
  representative photometric system covering the wavelength range $550
  \, \rm{nm} < \lambda < 1800 \, \rm{nm}$, and combine this
  information with different estimates for the evolution of the QSO
  luminosity function. We conclude that planned ground-based surveys
  such as Pan-STARRS and VISTA should be able to detect a large number
  of luminous QSOs up to $z \la 7.5$, but that space-based missions
  such as EUCLID (formerly SPACE/DUNE) or SNAP are probably required
  in order to obtain substantial samples at higher redshift. We also
  use our models to predict the expected number counts for future
  X-ray space missions (such as XEUS and Constellation-X), and show
  that because of their small field-of-view, these telescopes are
  unlikely to discover significant numbers of AGN at very high
  redshift. However, X-ray follow-up of objects detected at longer
  wavelength will be an important means of confirming their identity
  as AGN and constraining obscuration. 
\end{abstract}


\section{Introduction}\label{intro}

A number of observations point towards a tight relationship
between the properties of the AGN/QSO\footnote{We refer to bright AGNs
as QSOs, with no reference to their radio properties.}  population and
their host galaxies. This evidence includes the local relation between
the mass of the central supermassive black hole (SMBH) and the mass of
the spheroidal component of the host (e.g. H\"aring \& Rix, 2004), and
the apparent common ``downsizing'' behavior of star formation and
black hole accretion (Hasinger, Miyaji \& Schmidt, 2005). As well, it
is now becoming fairly widely accepted that the energy released by
accretion onto SMBH is an important mechanism in regulating galaxy
growth and star formation, although the details of how this ``AGN
feedback'' process works remain unclear.

In recent years, several theoretical studies have attempted to
understand the complex interplay between the physical mechanisms that
lead to the observed properties of galaxies and SMBH and their
evolution with redshift (for recent implementations see e.g., Monaco,
Fontanot \& Taffoni, 2007; Hopkins et al., 2007; Somerville et al.,
2007; Croton et al., 2006; Bower et al., 2006).  It has become evident
that the redshift evolution of the QSO population provides some of the
strongest constraints on this class of models (Fontanot et al., 2006;
Bromley, Somerville \& Fabian 2004). It is therefore of fundamental
importance to determine the statistical properties of high-redshift
QSOs, and in particular their luminosity function (LF), up to the
highest possible redshifts (Richards et al., 2006; Jiang et al., 2007;
Fontanot et al., 2007, hereafter F07). In addition, samples of high
redshift QSOs are important in order to place constraints on when and
how SMBH formed, and are of interest for studying the reionization
epoch (Gallerani et al. 2007).

A considerable difficulty in determining the high-z QSO LF is due to
the very low space-density of these objects, which are detected in
considerable numbers only in very large area surveys such as the Two
Degree Field QSO Redshift Survey (2QZ, Croom et al.  2004) and the
third edition of the Sloan Digital Sky Survey (SDSS) Quasar Catalog
(SDSSqso3, Schneider et al.  2005). As well, cosmological redshifting
and dimming, and the severe IGM absorption at such redshifts require
fairly deep Near-Infrared (NIR) observations. This combination of area
and depth in the NIR has not been achievable up until now.

However, a number of upcoming and proposed projects will begin to
change this situation. For example, several deep-wide NIR surveys from
the ground are in progress or planned, such as the
UKIDSS\footnote{http://www.ukidss.org/} Large Area Survey (LAS), Deep
Extragalactic Survey (DXS), and Ultra Deep Survey (UDS), the Panoramic
Survey Telescope and Rapid Response System
(Pan-STARRS)\footnote{http://www.ps1sc.org/index.htm} and
VISTA\footnote{http://www.eso.org/sci/observing/policies/PublicSurveys/
  \, sciencePublicSurveys.html\#VISTA} Ultra-VISTA, VIKING, VHS, and
VIDEO surveys.  Hopefully, Wide Field Camera 3 (WFC3) will soon be
installed on the Hubble Space Telescope, and with its 4.8 arcmin$^2$
Field of View (FOV), it will greatly improve our current abilities to
survey relatively large areas in the NIR from space. However, major
progress in characterizing the very high redshift QSO population will
probably have to wait for two kinds of future space telescopes. The
James Webb Space Telescope (JWST) will have a large aperture (6.5 m)
and high sensitivity ($\sim 3.5$ nJy), but its NIR imager NIRCAM will
have a relatively small FOV (2 $\times$ 4.7 arcmin$^2$). On the other
hand, several new space telescopes have been proposed with smaller
apertures ($\sim 1-2$ m) but with very large ($\sim 0.5$--1 sq. deg.)
optical and NIR cameras, such as SNAP\footnote{http://snap.lbl.gov/}
and EUCLID (formerly DUNE\footnote{http://www.dune-mission.net/} and
SPACE\footnote{http://www.spacesat.info/}). While the main motivation
of these latter kinds of missions is to constrain dark energy through
Supernovae, weak lensing, or baryon oscillations, as we will show in
this paper, they are also very well suited for the important goal of
studying the QSO population at very high redshift.

Complementary information on the evolution of the QSO population will
be provided by the next generation of X-ray space observatories.  Two
proposed missions are of particular interest in this regard:
XEUS\footnote{http://sci.esa.int/science-e/www/area/index.cfm?fareaid=103}
and Constellation-X\footnote{http://constellation.gsfc.nasa.gov/}. The
former is a 4.2 m telescope (minimum effective area 3 m$^2$ from 2 to
10KeV), with an imaging resolution better than 5 arcsec; the latter
consists of four coaligned 1.3 m telescopes on a single spacecraft
(minimum effective area 4 $\times$ 0.1 m$^2$), reaching 30 arcsec
resolution in the hard band.  Both have a relatively small FOV ($\sim
7$--10 and 5 arcmin$^2$, respectively).


The goal of this paper is to present predictions for the
observed-frame near-infrared colors and expected number density of
$z>5$ QSOs, which may help to guide the planning for future surveys
with these kinds of facilities. For comparison we also present
predictions in terms of the flux limit in the hard X-ray band. Of
particular interest is the trade-off between area and depth. Although
it is possible to make predictions for the relevant quantities using
physically motivated semi-analytic models set within the hierarchical
structure formation paradigm (e.g. Volonteri \& Rees 2006; Salvaterra
et al. 2007; Rhook et al. 2008), there are extremely large
uncertainties in these predictions at high redshift, associated with
our lack of knowledge about such factors as the nature of seed black
holes, the efficiency with which early black holes can grow, and
whether these holes can be ejected from their host galaxies by effects
such as the gravitational rocket (see e.g. Volonteri \& Rees 2006). We
therefore adopt a more empirical approach, in which we make use of
observed AGN luminosity functions at the highest available redshifts
($z\sim 5.5$--6) and template AGN spectral energy distributions
(SEDs). We combine these with an array of simple assumptions about how
the QSO population evolved with time to predict the observable
properties back to $z\sim 10$. In our simplest and most optimistic
model, we assume that the intrinsic properties of the underlying
population do not evolve at all. In the absence of strong evolution in
the SEDs of QSOs, this should be an upper limit to the number of
objects that can be detected at high redshift. In an intermediate
model, we extrapolate the observed evolution in the redshift interval
$3.5 < z < 5.2$ to higher redshift using a simple parameterization of
luminosity/density evolution developed by Fontanot et
al. (2007). Finally, in a semi-empirical model, we compute the
backwards evolution of the luminosity function assuming that QSOs at
$z\sim 6$ have grown by steadily accreting at their Eddington
luminosity. Although simple, these three models are empirically
motivated and probably provide a reasonable bracketing of the expected
results. As well as providing a good guess for the parameters of the
populations to be targetted by the future NIR and X-ray space missions
that we have mentioned, they will provide a useful foil for
predictions from more physically motivated formation models.

The structure of the paper is as follows. In section~\ref{clone} we
describe our technique for computing observed-frame QSO colors as a
function of redshift. In section~\ref{counts} we then combine this
information with different estimates for the evolution of the LF, in
order to bracket the expected QSO counts as a function of redshift,
area and depth. In section~\ref{x} we use a similar approach to
predict the corresponding quantities as a function of X-ray flux in
the $2-10 \, {\rm keV}$ band. Finally in Section~\ref{fin} we present
our conclusions.  Throughout this paper we assume a cosmology with
$h$, $\Omega_{\rm tot}$, $\Omega_m$, $\Omega_\Lambda$ = 0.7, 1.0, 0.3,
0.7; magnitudes and colors are in the AB system.

\section{Predicted QSO colors}\label{clone}

In order to estimate observed frame QSO colors and magnitudes at
$z>5$, we adopt the procedure described in F07. We consider a sample
of high-quality spectroscopic observations from SDSS, in the redshift
interval $2.2<z<2.25$, where the sample has the highest possible level
of completeness and, at the same time, the continuum of the QSOs is
sampled over the largest possible wavelength interval longwards of
$\mathit{Ly}_\alpha$. The final library consists of 215 SDSS spectra.
Then, for each object in the sample, we estimate the rest-frame
spectra, extending them up to near-infrared wavelengths using the
continuum fitting technique\footnote{They defined several “continuum
  windows” along the spectrum and fit the observed fluxes with a
  power law. The resulting best-fit parameters give an estimate of the
  intrinsic QSO spectrum blueward of the $\mathit{Ly}_\alpha$.}  of
Natali et al.  (1998). Blueward of the $\mathit{Ly}_\alpha$ a fixed
continuum slope has been assumed following Telfer et al.  (2002). The
final library is then used to clone QSO colors. In order to simulate
the spectra at different cosmic epochs, each template has been
redshifted and the IGM absorption computed by using the Madau, Haardt
\& Rees (1999) model, modified to match the Songaila (2004)
observations at $3<z<6$ (see the original F07 paper for more details
on the modified IGM model).  It is worth noting that this model
provides a deterministic prediction of IGM absorption as a function of
redshift.  At $z>6$ in particular the absorption is so severe that
almost no flux is predicted blueward of $\mathit{Ly}_\alpha$. The
variance in the IGM absorption along different lines of sight at fixed
redshift is expected to be important at such high optical depth (Fan
et al., 2006). The effect is particularly relevant for the tailoring
of high-completeness selection criteria based on color-color criteria,
but it is less important for the computation of the k-corrections and
the prediction of number counts, which are based on observed
magnitudes. Therefore in the following we will not consider a detailed
treatment of IGM statistics. We take advantage of the different
continuum slopes and strengths of the emission lines among the
template spectra to statistically estimate both the mean expected QSO
color and its variance as a function of redshift. In the original F07
paper the authors demonstrated that this procedure is able to recover
the colors of observed QSO in the SDSS photometric system up to
$z\sim5.2$.

\begin{figure}
  \centerline{
    \includegraphics[width=9cm]{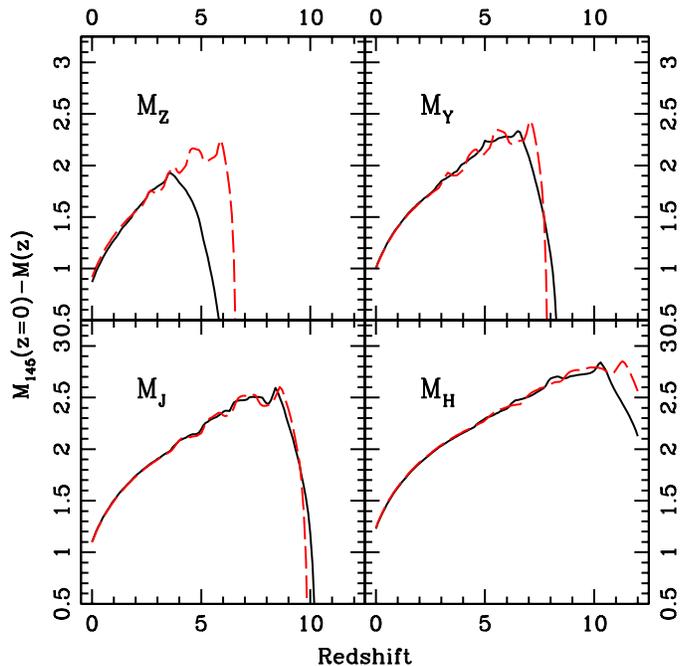}
  }
  \caption{Evolution of $M_{145}(z=0)-M(z)$ in the mock top-hat
    photometric system (solid lines) and in the UKIDSS photometric
    system (dashed lines).}
  \label{fig:kcorr}
\end{figure}

We explore a standard color-color selection technique for identifying
$z>5$ QSOs. To select objects in the redshift range of interest, we
require observations in roughly the $z$ through $H$ or $K$ bands. For
illustrative purposes, we consider a representative filter system
similar to the one being considered for the DUNE project.  This
consists of four top-hat filters: a broad optical filter covering the
$550$ to $920$ nm interval (roughly corresponding to the $r$, $i$ and
$z$ bands; in the following we refer to this filter as $Z$); a
$Y$-like filter from $920$ to $1146$ nm; a $J$-like filter from $1146$
to $1372$ nm; and an $H$-like filter from $1372$ to $1800$ nm. This
system has the advantage of perfect coverage of the wavelength
interval of interest at $5<z<12$, and it gives a good representation
of the real filter configurations being considered by the projects we
have discussed. In fig.~\ref{fig:kcorr} we show the predicted redshift
evolution of the difference between the absolute restframe magnitude
at 145 nm ($M_{145}$) and the absolute magnitude corresponding to the
four filters. We chose $M_{145}$ as a reference magnitude to compare
our prediction to the estimate of the QSO LF given in F07. We note
that the $\mathit{Ly}_\alpha$ line exits each of our four filters at
$z=6.57$, $z=8.42$, $z=10.28$ and $z=13.80$ respectively. 
Given the strong IGM $Ly_\alpha$ resonance absorption at $z>6$, for
each band these correspond to the redshifts at which the QSOs become
``drop-outs''. In the same figure we also plot the predictions of our
approach for a more realistic photometric system, namely the UKIDSS
ZYJH filter system (dashed line). It is evident from the figure that
the response of two photometric systems are in good agreement in the
redshift range of interest.

\begin{figure}
  \centerline{
    \includegraphics[width=9cm]{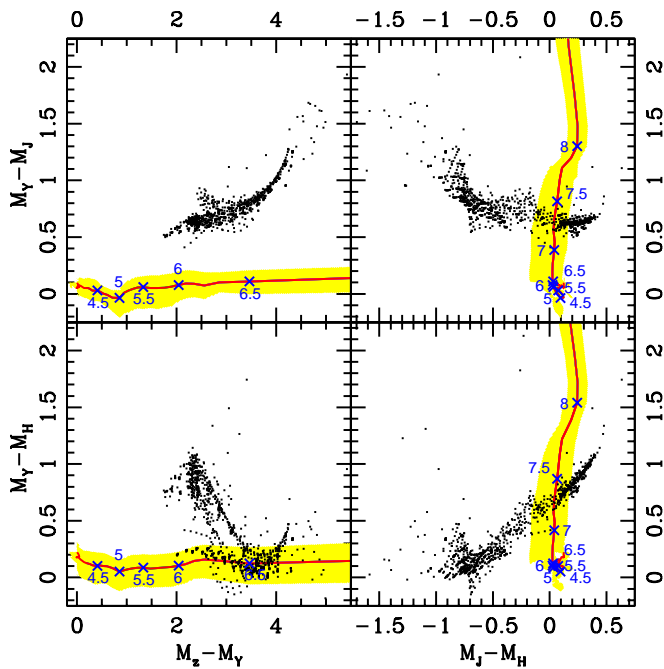}
  }
  \caption{QSO color evolution as a function of redshift in the mock
    top-hat photometric system. The shaded region represent the
    scatter in the template library ($5\%$ and $95\%$ percentiles of
    the distribution). Dots mark the expected colors of brown dwarfs.}
  \label{fig:colo}
\end{figure}

We show in fig.~\ref{fig:colo} four sections of the four dimensional
color spaces. The most numerous astrophysical contaminants in dropout
searches for high-redshift quasars are cool, low-mass stars (brown
dwarfs). The surface density of brown dwarfs at the extremely faint
magnitudes considered here is completely unknown and cannot be
extrapolated from existing brown-dwarf surveys, though an upper limit
is given by the number of dwarf stars in the Milky Way (Wolfgang
Brandner, priv.  comm.). This is because faint brown dwarfs are the
oldest, coolest objects, and their exact magnitude and color
distribution depends on the details of the cooling process as well as
the distribution of ages and initial masses. For the same reason, it
is impossible to quantify the likely number of large-sigma outliers
from the mean brown-dwarf locus that would be scattered into the
quasar locus.  Hence, any brown dwarf sample discovered as by-product
of a quasar search would be extremely valuable to the brown-dwarf
field. A large fraction of them would be distinguishable from quasars
by the $H-K$ color, which is very blue in cold brown dwarfs due to
their much higher opacity in the $K$-band than in $H$. We therefore
compare our color evolution with the expected colors for brown dwarfs,
computed using both the spectral library defined by Reid et al.
(2001), and the theoretical spectral model from Burrows et al.
(2003,2006) and Hubeny \& Burrows (2007). It is evident from
fig.~\ref{fig:colo} that it is possible to define suitable color
criteria in order to largely disentangle the two populations.
However, since our filter system is only representative, we do not
attempt to define quantitative color criteria here. We compare our
predictions with simulations and observed colors of quasars in the
SDSS and UKIDSS systems (see Chiu et al., 2005, 2007; Hewett et al.,
2006); a comparable paper for Pan-STARRS is in preparation (S.
Jester, et al. in prep). The good agreement of fig.~\ref{fig:colo}
with, i.e., fig.~11 in Chiu et al.  (2005) provides additional
evidence that our results are representative, despite the idealized
photometric system we assume.  We also considered a more standard
(narrower) $z$-band instead of the broad-band $Z$ filter. We find that
the predicted locus of QSO candidates (yellow shaded region) is not
affected by the change; however the narrower filter shows a faster
color evolution with respect to fig.~\ref{fig:colo}. The difference is
due to the larger $\mathit{Ly}_\alpha$ forest flux in the broad-band
filter at $z<6$.  We therefore conclude that the use of a broader
filter allows for a more reliable photometric redshift estimate.

\section{Expected QSO counts} 
\label{counts} 

\begin{figure*}
  \centerline{ \includegraphics[width=9cm]{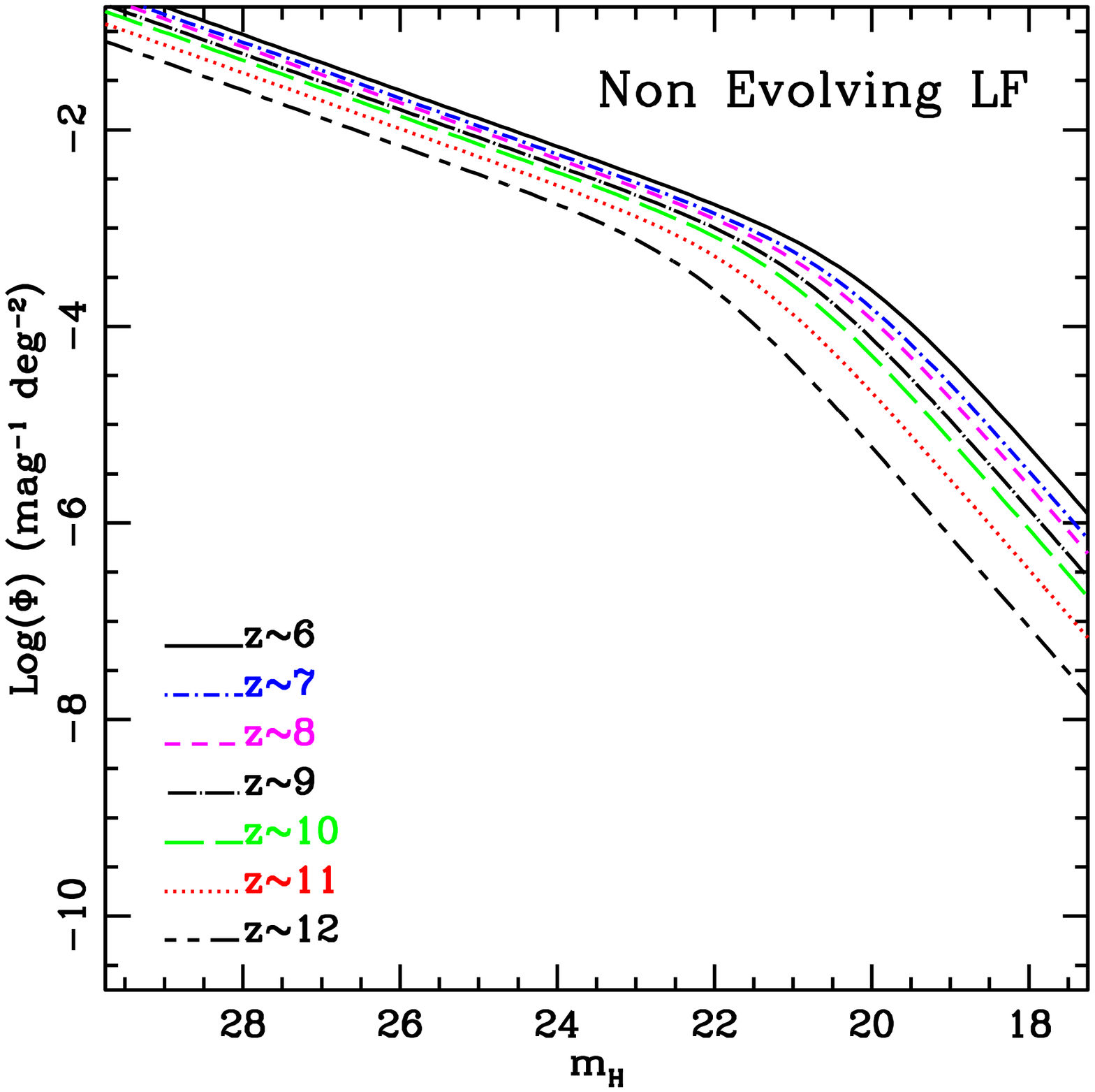}
    \includegraphics[width=9cm]{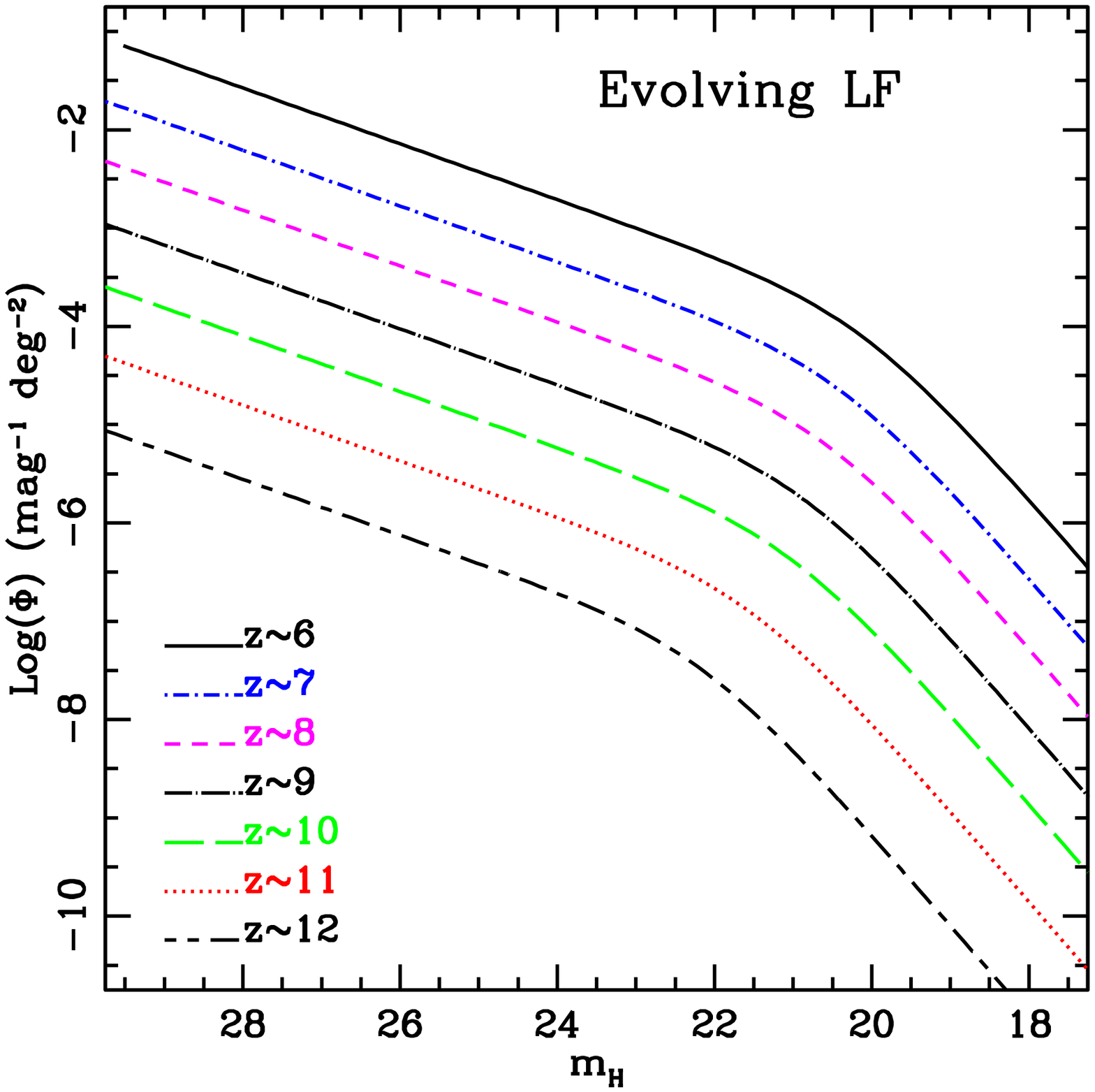} }
  \caption{Expected QSO LF evolution in apparent $m_H$ magnitude at
    different redshifts. The left panel shows the non-evolving LF
    prescription, the right panel the evolving LF prescription, based
    on the best-fit F07 LF (Nr.13a; see text).}
  \label{fig:lf}
\end{figure*}

\begin{figure}
  \centerline{ \includegraphics[width=9cm]{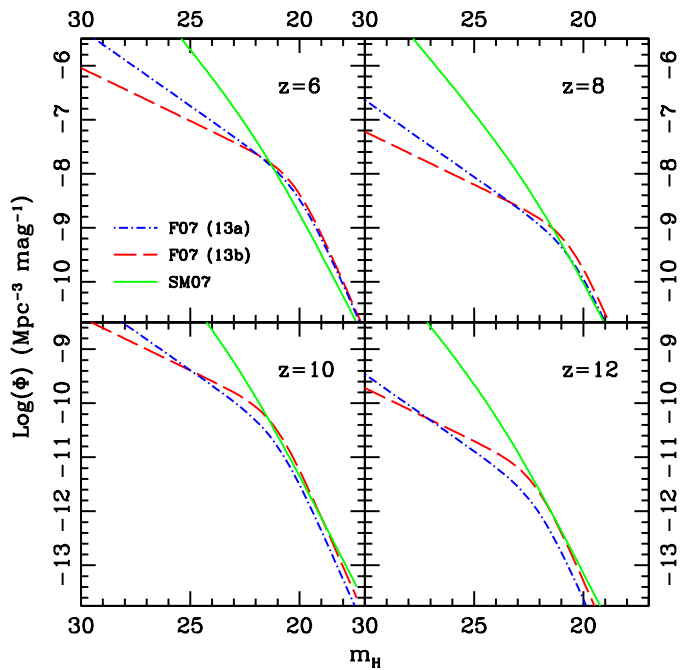} }
  \caption{Comparison between the LFs in apparent $m_H$ magnitude
    predicted by the model of F07 (dot-dashed and dashed line refer
    respectively to model Nr. 13a and Nr. 13b respectively) and SM07
    (solid line) models at different redshifts.}
  \label{fig:clf}
\end{figure}

\subsection{Models for QSO Evolution}
Combining our estimate of $M_{145}-m_H$ with an estimate of the QSO LF
at $z>5$, we can now compute the expected QSO counts in a hypothetical
survey as a function of redshift, area and $H$-band depth. We will
consider three different models for the evolution of the intrinsic
properties of the QSO population: non-evolving luminosity function,
evolving luminosity function, and Eddington-limited accretion. In the
first model (non-evolving LF), the comoving space density of QSOs is
kept fixed to the observed value given by the adopted $z=5$ LF, so
that the evolution in the observed population is entirely driven by
dimming due to the growing luminosity distance, plus k-corrections. In
the second class of models, we make use of the analysis of F07, in
which they combined a faint QSO sample obtained from the GOODS survey
(Giavalisco et al., 2004), with a bright QSO sample extracted from the
SDSSqso3, in order to study the QSO LF and its evolution in the
redshift interval $3.5<z<5.2$.  They found that Pure Density Evolution
models were a better representation of the observed QSO population at
these redshifts than Pure Luminosity Evolution models. Their best fit
model (Nr. 13a in the original paper) requires an evolution of the
magnitude of the knee of the double power law, a relatively steep
faint-end slope and a bright-end slope as steep as local observations
(in disagreement with Richards et al., 2006; for a discussion of the
slope estimates we refer the reader to F07). However, F07 also found a
good match between observations and the prediction of a similar model,
with a shallower faint-end slope (their model Nr. 13b). These two
models roughly correspond to the faint-end slope estimates at lower
redshifts given by Richards et al. (2005). Given the importance of the
faint-end slope for the prediction of QSO number counts in small area
deep surveys, we therefore consider both models in the following
discussion.

In fig.~\ref{fig:lf}, we show our assumed LFs, as a function of
apparent $m_H$, based on the best-fit F07 model (13a), for both the
non-evolving and evolving LF. We compare the F07 results with the
recent analysis of the QSO-LF at $5<z<6.5$ proposed by Shankar \&
Mathur (2007, hereafter SM07). They have re-analyzed the SDSS bright
sample at $z>5$ (Fan et al., 2001,2004), in the light of recent
observations of faint QSOs (Willott et al.  2005; Cool et
al. 2006). Their LF has the same double power-law functional form as
in F07, while Fan et al.  (2004) used a single power law; in order to
correctly reproduce the redshift evolution of the SDSS bright sample,
SM07 renormalized their LF to the cumulative number density given in
Fan et al. (2004) at $z=6.07$, and at each redshift they required the
bright-end to match the Fan et al.  (2004) LF at $M_{145}=-27.00$
(Shankar, private communication). The SM07 LF has a redshift
evolution\footnote{such as $\Phi(z) = \Phi_{(z=6.07)}
  10^{-0.48(z-6.07)}$} as in Fan et al.  (2001), and a steeper
faint-end slope ($<-2.0$) with respect to the F07 results.  In the
following we consider the 90\% CL faint-end slope for the optical LF
given in SM07. In fig.~\ref{fig:clf} we compare the predictions for
this model and the two F07 models at different redshift. It is worth
noting here that the two groups worked at different redshifts
($3.5<z<5.2$ for F07 and $5.0<z<6.5$ for SM07) and that we are
extrapolating their results well beyond these confidence
intervals. Keeping these caveats in mind, the overall agreement is
quite good.

We compute the number of expected QSO per unit area by integrating the
resulting LF up to a limiting magnitude. In order to account for the
uncertainties in the evolution of the LF we apply a bootstrap
technique. For both F07 models, the redshift evolution is quantified
using an exponential form\footnote{such as $\Phi(z) = \Phi_{(z=2)}
  e^{k_z((1+z)-3)}$}: the authors also give an estimate for the error
on this parameter based on their minimization algorithm.  Here we
consider 1000 Monte Carlo realizations of the expected QSO counts,
randomly varying the value of $k_z$ over the range of the quoted
error. We repeat the procedure for both models for the evolving
LF/non-evolving LF and we use the results (mean and variance) to
define our predicted range in number counts.

In our third class of models (Eddington), we compute the LF evolution
under the hypothesis that the SMBHs responsible for the observed LF at
$z\sim 6$ have accreted at the Eddington rate during their entire past
history. This gives an exponential mass evolution (see i.e. Volenteri
\& Rees, 2006):

\begin{equation}
  M_{BH}(t) = M_{BH}(0) \exp \big( \frac{1-\eta}{\eta} \frac{t}{t_{Edd}} \big )
\end{equation}

\noindent 
where $t_{Edd}=0.45$ Gyr and $\eta =0.1$ is the radiative efficiency.
We convert the mass evolution to luminosity evolution and we predict
the corresponding number counts as a function of redshift and depth.
We consider both F07 parameterizations (13a and 13b) of the LF as well
as the SM07 results and we fixed $z=6$ as the initial redshift for the
accretion at the Eddington rate.

\subsection{Results}

\begin{figure*}
  \centerline{
    \includegraphics[width=9cm]{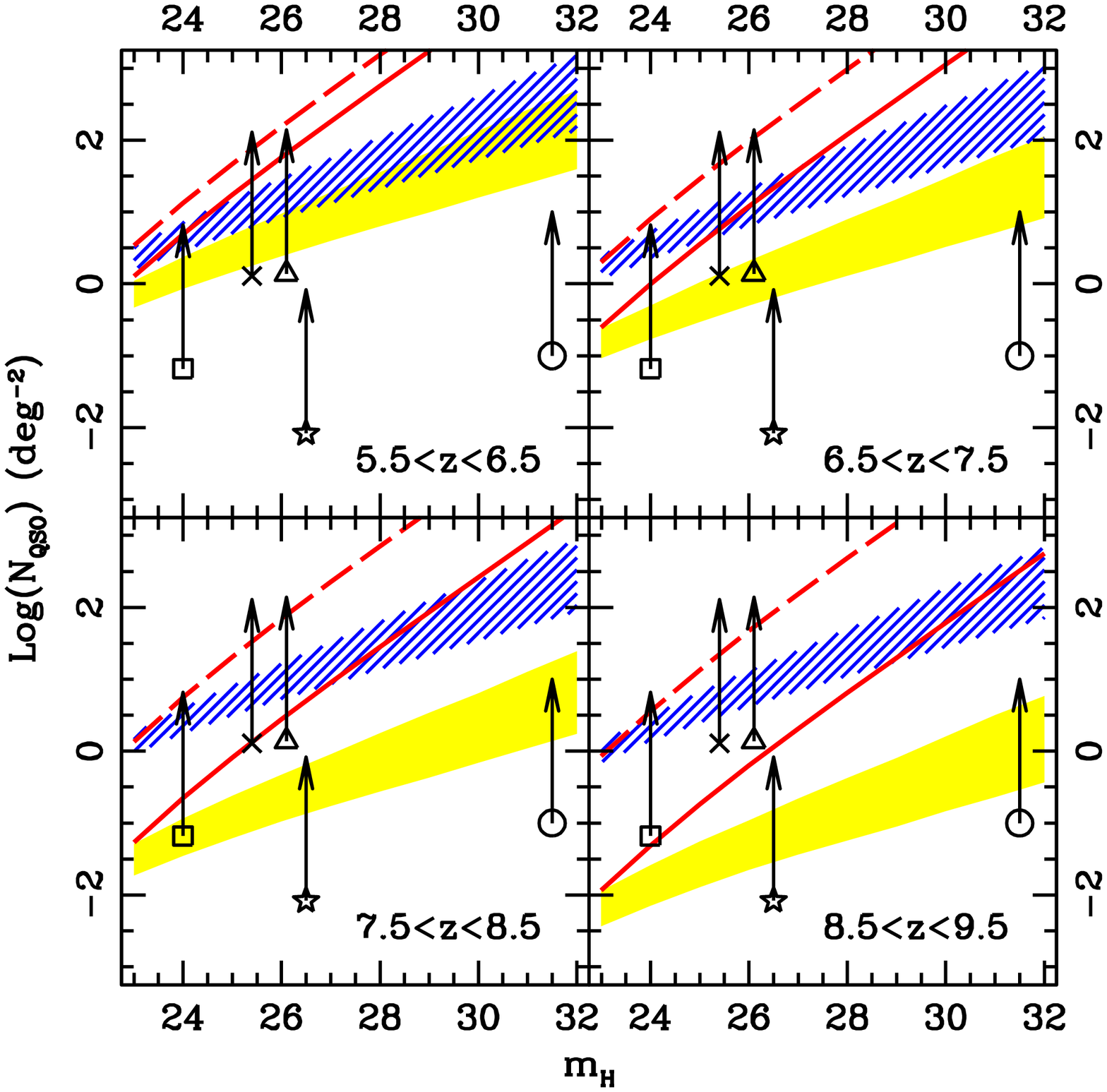}
    \includegraphics[width=9cm]{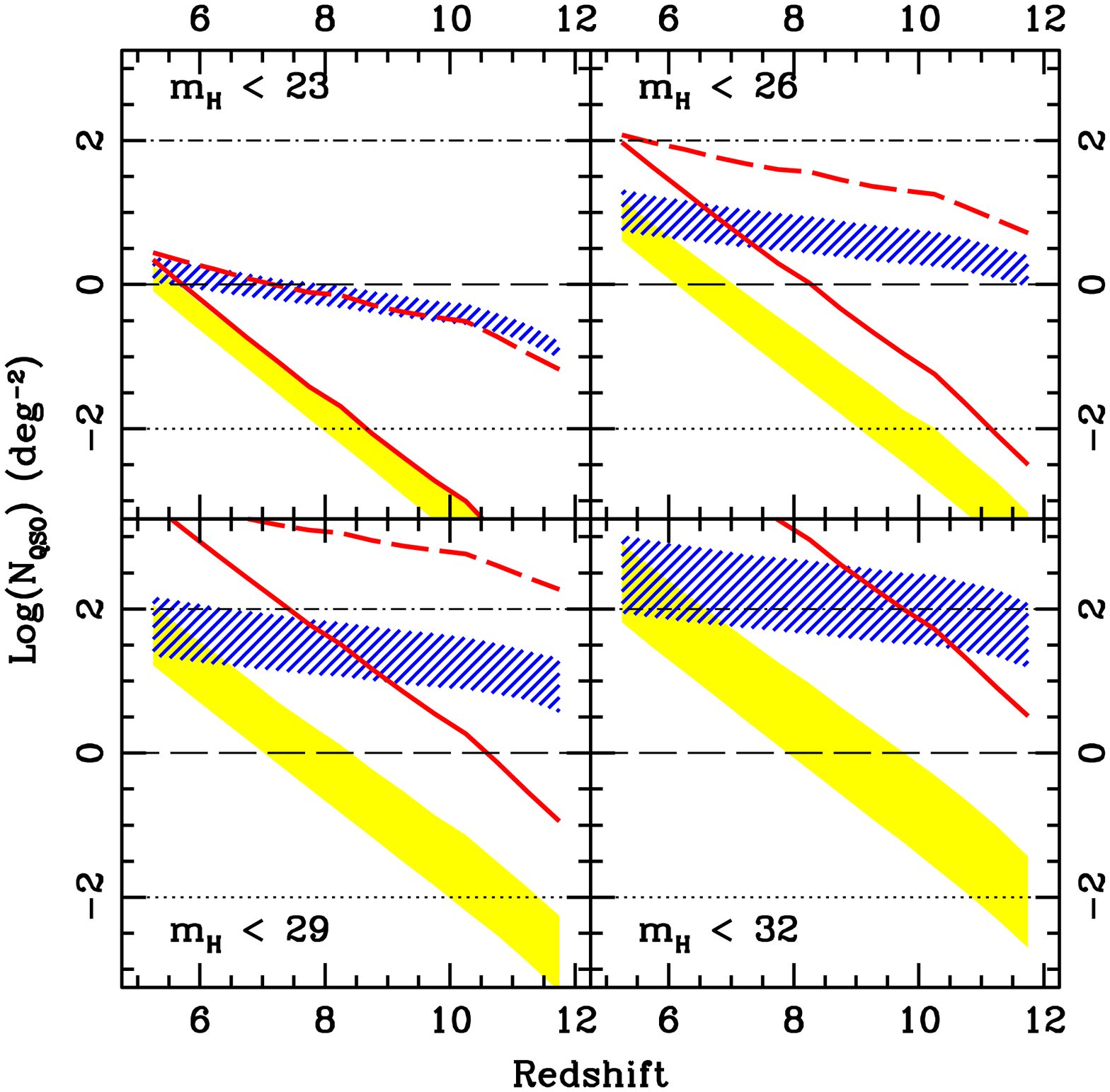}
  }
  \caption{In both panels, the diagonal hatched area shows the
    predictions of the non-evolving LF model, and the shaded area
    shows the evolving LF model, both based on the best-fit F07
    LF. The shaded regions are estimated via Monte Carlo simulations
    accounting for the errors in the LF parameters. The solid/dashed
    line shows the predictions for the evolving/non-evolving LF
    models based on the SM07 LF.
    Left Panel: QSO counts as a function of limiting magnitude.  The
    symbols mark the surface density (see also tab.~\ref{tab1})
    corresponding to one detected QSO in UDS (cross), Ultra-VISTA
    (triangle), VISTA VIDEO (square), 
    DUNE (star), JWST (circle), while the tip of the arrow shows the
    surface density that would result in the detection of at least 100
    QSOs. Right Panel: QSO counts as a function of redshift. The thin
    dotted, dashed and dot-dashed horizontal lines mark the levels
    corresponding to $0.01$, $1$ and $100$ QSO per square degree
    respectively. }
  \label{fig:nc2}
\end{figure*}

Fig.~\ref{fig:nc2} (left panel) shows the predicted number of QSOs per
square degree as a function of limiting magnitude for four redshift
intervals ($5.5<z<6.5$, $6.5<z<7.5$, $7.5<z<8.5$, $8.5<z<9.5$). We
also show the same quantity as a function of redshift in the right
panel. The four panels refer to different magnitude depths ($23.0$,
$26.0$, $29.0$, $32.0$). As a reference a $m_H=23.00$ magnitude limit
corresponds to absolute magnitudes $M_{145}=-23.46$, $-23.73$,
$-23.90$ and $-24.16$ at $z=6$, $7$, $8$ and $9$ respectively. We show
both the non-evolving and evolving versions of models based on both
the F07 LF and the steeper SM07 LF. The difference (several orders of
magnitude) at faint magnitudes between the F07-based and the
SM07-based models underlines the large uncertainties associated with
the faint end of the AGN LF even at lower redshifts, which naturally
blow up as we extrapolate these results to higher redshift.

\begin{table*}
  \begin{center}
    \begin{tabular}{cccc}
      \hline
      Survey & Depth ($m_H$) &  Area ($deg^{2}$) & \\
      \hline
      UDS & 25.4 & 0.77 & \\
      Pan-STARRS Medium Deep Survey & $y<24.8$ & 80 & \\
      Pan-STARRS $3\pi$ Survey (extragalactic part) & $y<21.5$ & 20,000 & \\
      Ultra-VISTA & 26.1 & 0.73 & \\
      VISTA VIDEO & 24.0 & 15 & \\
      SPACE All Sky Survey & 23.0 & 30,000 & \\
      SPACE Deep Survey & 26.0 & 10 & \\
      DUNE Medium-Deep & 26.5 & 120 & \\
      JWST & 31.5 & 10 & \\
      \hline
      & Point Source sensitivity &  Field of View & \\
      & (for 1 Ms observation) & (${\rm arcmin}^{2}$) \\
      \hline
      XEUS & $3 \times 10^{-18} \, {\rm erg/s/cm^{-2}}$ & 7 \\
      Constellation-X & $2 \times 10^{-17} \, {\rm erg/s/cm^{-2}}$ & 5 \\
      \hline
    \end{tabular}
    \caption{Planned depth and area for the surveys mentioned in the text.}
    \label{tab1}
  \end{center}
\end{table*}

In fig.~\ref{fig:nc2}, the symbols show the surface density that would
result in the detection of at least one QSO in the UDS, Ultra-Vista,
and VISTA-VIDEO surveys at the corresponding $H$-band magnitude depth
of the survey, while the tip of the arrow shows the surface density
that would result in the detection of at least 100 QSOs. In addition
to these already-approved ground-based surveys, we show space-based
surveys that could be carried out with the EUCLID and JWST missions.
We collect the information about the depth and area of each survey in
table~\ref{tab1}, where we also list the $y$-band depths of the
Pan-STARRS $3\pi$ and Medium Deep Surveys; however since those surveys
do not include an $H$-band filter, we do not show them in the plot.
For a given magnitude limit and area, the shaded swaths or lines
should lie above the arrow tip in order to obtain $>$100 QSOs, or
above the symbol in order to detect at least one QSO. One should keep
in mind that the diagonal-hatched swaths and dashed lines assume no
evolution in the underlying QSO number densities, and are probably
overly optimistic. If we take our best-fit F07-based evolving LF model
as representative, we see that both the JWST and EUCLID-like surveys
are expected to detect a few hundred QSOs in the redshift ranges $6.5
< z < 7.5$ and $6.5 < z < 7.5$, and a few tens at $8.5<z<9.5$.
However, JWST will mainly constrain the faint end of the QSO LF, while
wide-field missions like EUCLID or SNAP are needed to detect
significant numbers of bright QSOs.

\begin{figure*}
  \centerline{
    \includegraphics[width=9cm]{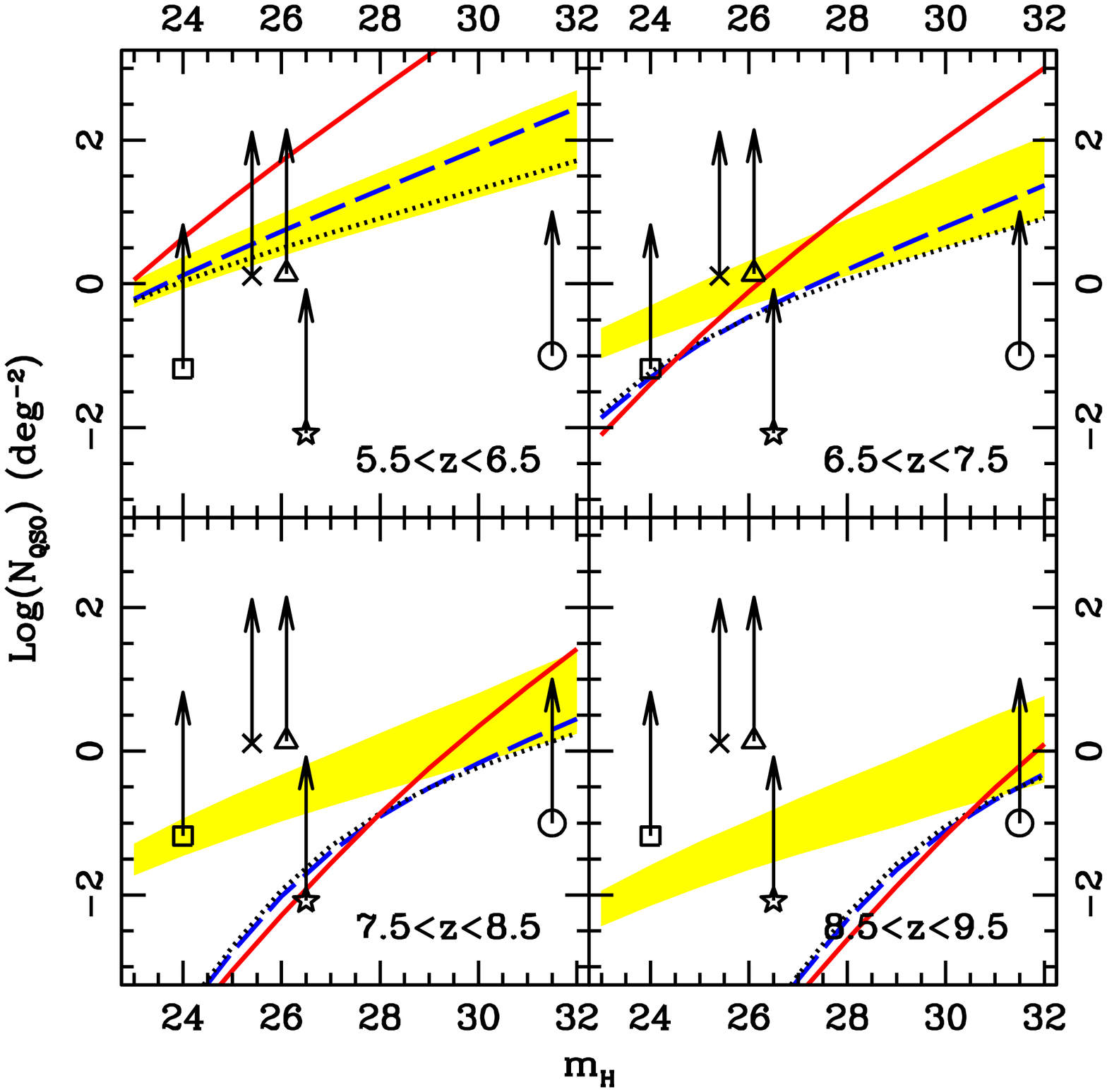}
    \includegraphics[width=9cm]{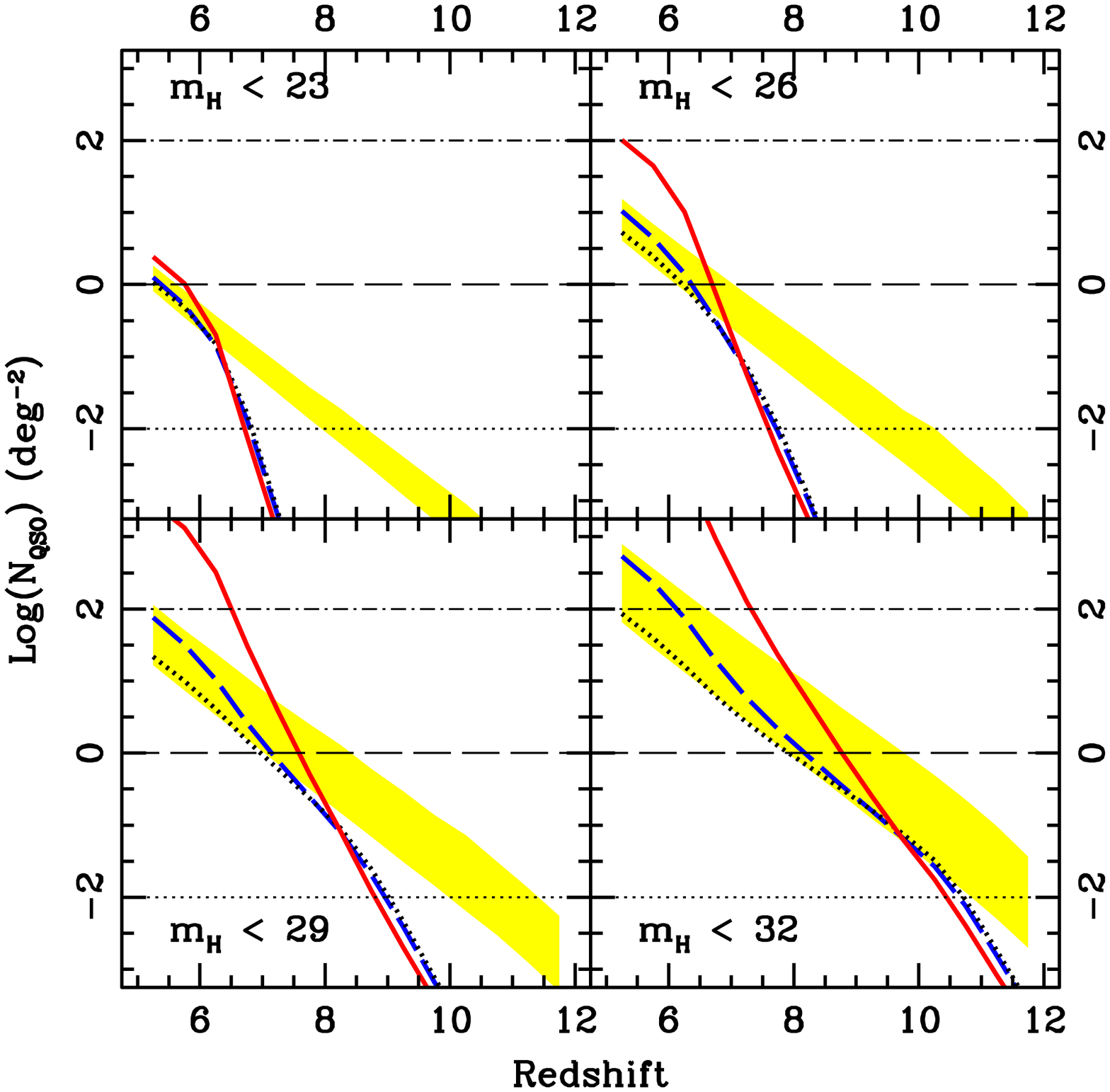}
  }
  \caption{In both panels, the solid, dashed and dotted lines show the
    predictions for the Eddington accretion models calibrated to the
    SM07 and F07 (model Nr.13a and Nr.13b) LFs respectively. For
    comparison, the shaded region reproduces the predictions of the
    F07-based evolving LF model (as shown in Fig.~\ref{fig:nc2}).
    Symbols and thin lines are as in fig.~\ref{fig:nc2}. Left Panel:
    QSO counts as a function of limiting magnitude.  Right Panel: QSO
    counts as a function of redshift.}
  \label{fig:nc2t}
\end{figure*}

Fig.~\ref{fig:nc2t} shows the prediction of the Eddington accretion
model when calibrated to the SM07 and F07 (model Nr.13a and Nr.13b)
LFs (see figure caption for model key). To guide the eye, in the same
plots we repeat the predictions for the evolving F07 LFs as a shaded
area. We can see that the Eddington model predicts much more rapid
evolution in the numbers of luminous QSOs, while the predicted numbers
of faint objects are within the shaded area of the F07 evolving LF
predictions. This illustrates the importance of probing the luminous
QSO population for constraining models of BH growth and evolution.

We summarize the whole set of predictions for the empirical and
semi-empirical models in tab.~\ref{tab:counts1} and
tab.~\ref{tab:counts2} respectively.

\section{X-ray Surveys}\label{x} 

\begin{figure*}
  \centerline{
    \includegraphics[width=9cm]{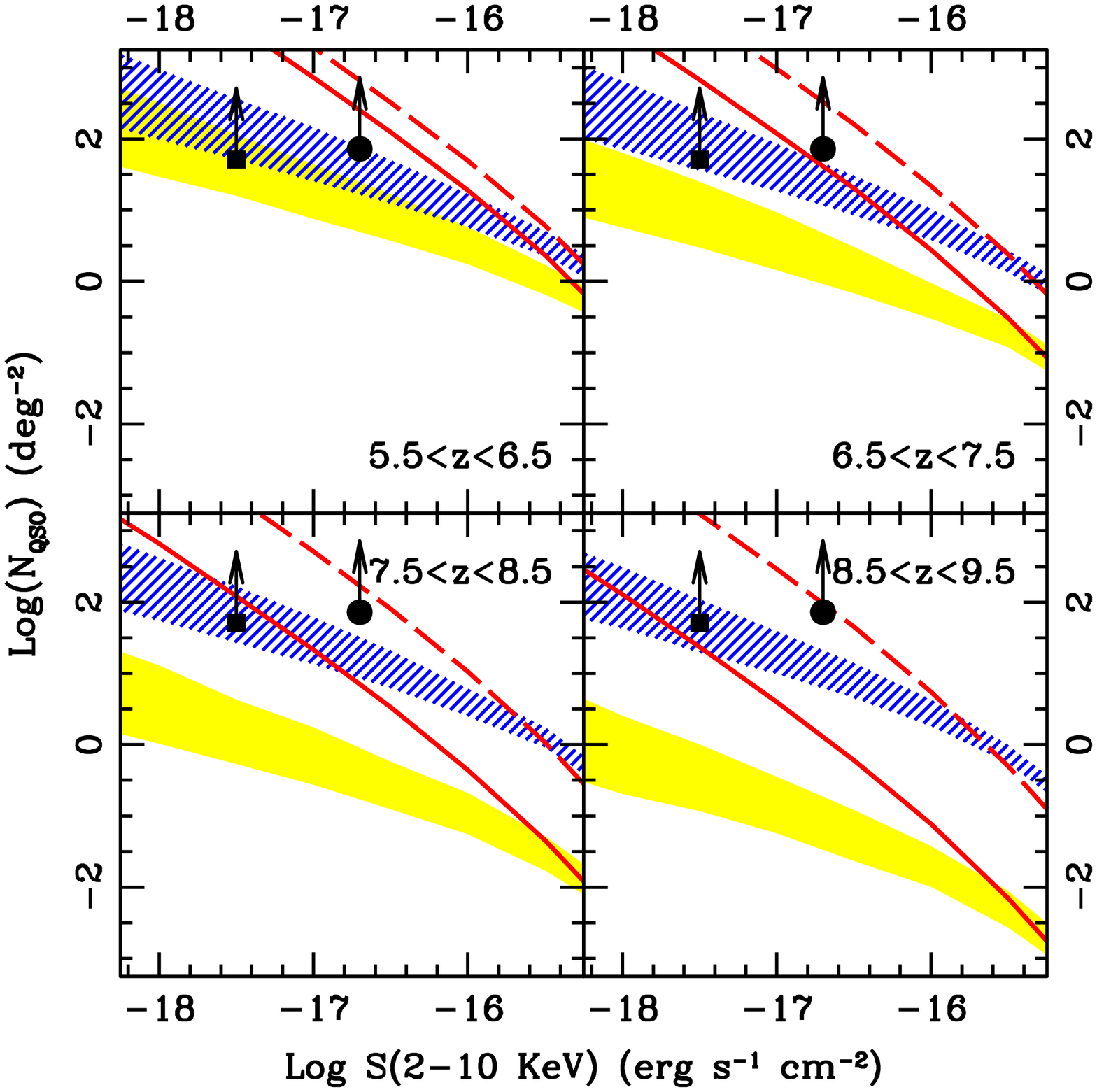}
    \includegraphics[width=9cm]{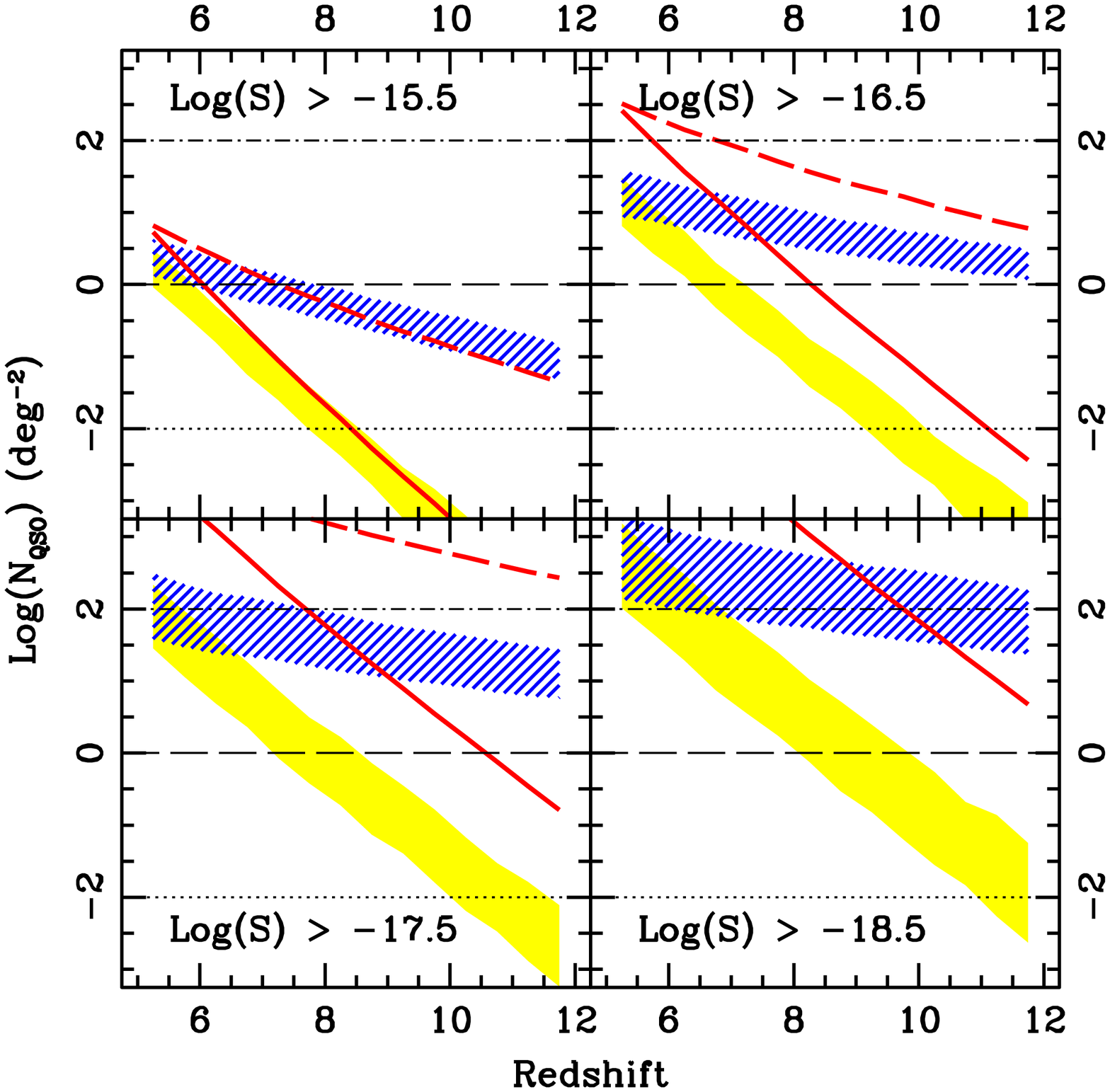}
  }
  \caption{Predicted QSO counts in the X-ray 2-10 keV band, for
    non-evolving and evolving LF models as in Fig.~\ref{fig:nc2}.
    Left Panel: QSO counts as a function of limiting X-ray flux. The
    symbols mark the expected point source sensitivity and inverse
    survey area for Constellation-X and XEUS (tab.~\ref{tab1}). Right
    Panel: QSO counts as a function of redshift. We see from this
    figure that blank sky X-ray surveys with these proposed missions
    are unlikely to be an effective means of discovering very high
    redshift AGN. }
  \label{fig:xnc2}
\end{figure*}

We can use the same approach to estimate the expected number of QSO
detected in the hard ($2-10$ keV) X-ray band as a function of
redshift, area and flux. We convert the $m_{145}$ magnitudes into
X-ray fluxes using the same approach as Fontanot et al. (2006): we
compute the bolometric LF using a restframe band correction between
$145$ nm and the B-band and the bolometric correction of Elvis et al.
(1994); we then convert bolometric into 2-10 keV luminosities
following Marconi et al., (2004), and we apply a k-correction in the
hard band, calibrated assuming a spectrum with photon index $\Gamma =
-1.8$. We do not attempt to correct for the luminosity-dependent
fraction of obscured and compton-thick objects. We show the predicted
X-ray number counts in fig.~\ref{fig:xnc2} 
corresponding to the same LF models we introduced in~\ref{counts}.
In a 1 Ms exposure, XEUS and Con-X are expected to be able to reach a
point source sensitivity of $3 \times 10^{-18}$ and $2 \times 10^{-17}
\, {\rm erg/s/cm^2}$, respectively (Hasinger et al., 2006). Assuming a
maximum total exposure time of 10 Ms, the largest area survey possible
with these missions will be about 100 arcmin$^2$ (or about $2.8 \times
10^{-3}$ sq. deg.). We show the area/depth markings for XEUS and Con-X
in fig.~\ref{fig:xnc2}, as before, except that we show only the symbol
that indicates ten AGN per field because the arrow tip would be off
the top of the plot. From this we see that blank sky X-ray surveys
even with this next generation of missions are unlikely to be an
effective means of discovering very high redshift AGN. They may find a
few tens to a hundred AGN at $z\sim 6$, but they are unlikely to
detect any objects at higher redshift.


\section{Discussion and Conclusions} \label{fin}

In this paper we present empirical predictions of the observed colors
and number densities of very high redshift ($z>5$) QSOs. We combine a
representative photometric system with a QSO spectral template library
to predict the evolution of QSO colors at $z>5$. Using a set of four
optical-NIR filters, we show that it is possible to define color-color
criteria that select high-z QSO candidates on the basis of their
photometric properties. We then combine the estimated k-corrections
with different models for the LF evolution in order to estimate the
expected number of QSOs as a function of surveyed area, magnitude
limit, and redshift.

We confront our findings with the parameters of existing, planned
and proposed surveys.  First, the SDSS has already covered the bright
end of the LF to $M_{145}\approx-26.5$ at $z<6.4$, and is pursuing
efforts to push another 2 magnitudes fainter over the 300 square
degrees of its deep southern stripe (Jiang et al.\ 2007).  The
Pan-STARRS survey, which is about to begin data-taking, will reach $z$
magnitudes that are up to 1\,mag fainter still over 20,000 square
degrees of extragalactic sky in its $3\pi$ survey, putting of order
3000 low-luminosity quasars at $6< z<6.4$ within its reach (from an
extrapolation of the Shankar \& Mathur LF).  Since the $3\pi$ survey
includes deep $z$ and $y$ band imaging, it is also ideally suited for
the discovery of quasars at redshifts $6.6 \le z \le 7.3$. Again
integrating the Shankar \& Mathur LF to the $y$-band flux limit, we
expect that Pan-STARRS will contain of order 150 quasars at $6.6 \le z
\le 7.3$.

In this paper we address the prospects for finding fainter quasars
than are accessible with SDSS and Pan-STARRS at $5<z<7.5$, or
\emph{any} quasars at $z>7.5$, with currently planned or proposed
surveys. It is evident from fig.~\ref{fig:nc2} that up to $z\la 6.5$,
the ground-based near-infrared UKIDSS and VISTA surveys should be able
to produce samples of roughly tens to hundreds of low-luminosity QSOs,
depending on the actual evolution of the LF. Finding quasars in the
Pan-STARRS Medium Deep Survey (MDS) requires complementary
near-infrared imaging (ideally in $J$) at the depth of VIDEO (3 of
whose fields are included in 3 of the MDS fields); if such imaging is
available, the larger area would make about 5-6 times more quasars
accessible than VIDEO alone.

\begin{figure}
  \centerline{
    \includegraphics[width=9cm]{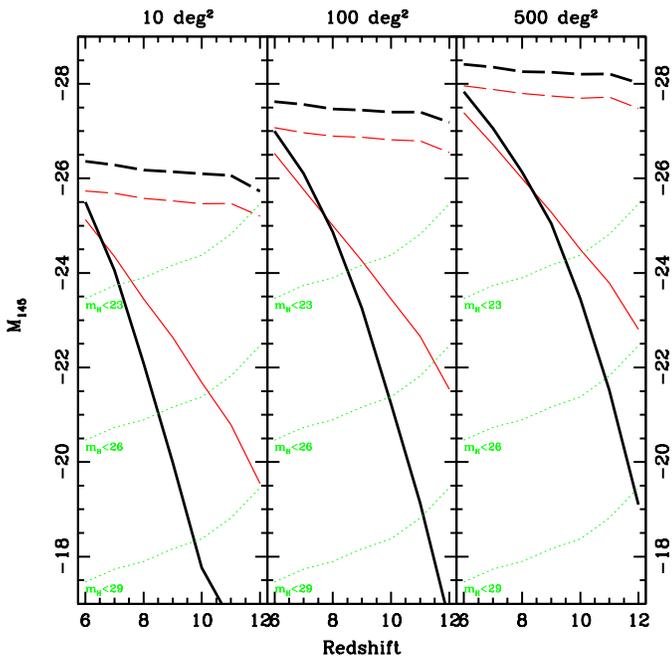}
  }
  \caption{The magnitude of the brightest QSO in a surveyed area of 10
    deg$^2$ (left), 100 deg$^2$ (middle), 500 deg$^2$ (right). Solid
    and dashed lines correspond to the evolving LF and non-evolving LF
    prescription respectively. Thick and thin lines correspond to the
    F07a and SM07 LF respectively. Dotted lines show the faintest
    absolute limiting magnitude as a function of redshift for a
    magnitude-limited sample.}
  \label{fig:newplot}
\end{figure}

Even in the absence of a decline in the space density of QSOs,
however, our results suggest that in order to obtain samples of
substantially more than a hundred low-luminosity QSOs at $z\ga 6.5$,
we will require the larger survey volumes that are accessible only
from space (because of the much lower NIR background, a 2m telescope
in space is more efficient at surveying large areas than an 8--10m on
the ground). It is also possible to use our predictions to estimate
the magnitude of the brightest QSO we expect, as a function of
redshift, at a fixed surveyed area (fig.~\ref{fig:newplot}). We define
this quantity as the magnitude corresponding to an integrated space
density of 1 QSO over the surveyed volume (in order to compute the
volumes we consider the same redshift intervals as in
fig.~\ref{fig:nc2}, left panel).  We consider the evolving and
non-evolving prescriptions (solid and dashed line respectively)
applied to the F07a and SM07 LFs (thick and thin line respectively).
In the same plot we also show the faintest magnitude reached at a
given depth in the $H$-band (dotted lines). Combining the two
informations, we are then able to define an accessible magnitude
interval.
Our results highlight that, while JWST will be able to detect
extremely faint QSOs, because of its small FOV it is unlikely that it
will be possible to survey large enough areas to detect significant
numbers of the most luminous QSOs at very high redshift. For this
important goal, we need the wider-area surveys that could be carried
out by missions like EUCLID or SNAP.  Similar arguments hold if we
consider planned X-ray surveys with missions such as XEUS and
Constellation-X. The very small FOV of these surveys, combined with
the relatively long exposure times needed to reach high enough
sensitivity, probably preclude the possibility of detecting large
samples of $z>6$ QSOs using the X-ray information alone. Targetted
X-ray observations of objects detected in the NIR will, however, be
crucial for confirming their identity as AGN and constraining
obscuration.

The predictions presented here are complementary to those based on
semi-analytic models set within the hierarchical structure formation
paradigm. For example, Salvaterra, Haardt \& Volonteri (2007) have
presented predictions for the properties of the AGN population that
could be observed in deep X-ray surveys with future surveys, based on
the Volonteri \& Rees (2006) models with different assumptions.  They
consider a merger tree based model based on the Press-Schechter
formalism and Eddington accretion onto the SMBH. An alternative model
has been presented by Rhook \& Haehnelt (2008). Their hybrid models
also assume that QSO activity is triggered by major merger events, and
they combine this information with different sets of assumptions for
the decline of the accretion rate onto the SMBH.  
As in our empirical models, their predicted counts span several orders
of magnitude. This underlines the importance of the observational
determination of the statistics of the QSO population in order to put
constraints on models of SMBH growth. We plan to compare the empirical
results presented here with additional semi-analytic, merger-tree
based models for SMBH evolution (Fontanot et al. 2007; Somerville et
al. 2008) in a forthcoming paper.

We expect that only for the brightest objects will spectroscopic
follow-up be feasible. The development of adequate photometric
redshift estimators, as well as a larger multi-wavelength coverage, is
therefore fundamental for the study of the faint population. Moreover
a simple drop-out technique is not able to disentangle a priori faint
QSOs from Lyman Break Galaxies at comparable redshift.  A key goal in
the study of faint high-z QSOs is therefore the definition of reliable
criteria to separate QSOs from galaxies. Here, we simply assume that
we can perfectly disentangle the faint QSOs from the contaminants such
as galaxies and stars. Under this hypothesis we consider a simple test
to assess the impact of the uncertainties in the photometric redshift
determination on the overall estimate of the QSO-LF and its space
density evolution.  We consider the F07 Monte Carlo algorithm for the
estimate of the LF, which combines the GOODS and SDSS samples. We then
introduce a random relative error in the redshifts of the faint
(GOODS) objects, perform 1000 bootstrap realizations of LF fitting,
and consider the new estimate of the LF parameters. We conclude that
the F07 technique is still able to recover the statistical properties
of the LF in the presence of these photometric redshift errors:
assuming a $5\%$ ($10\%$) error for the photometric redshifts, we add
only $0.3\%$ ($0.5\%$) to the error on the redshift evolution and
$1.0\%$ ($1.5\%$) to the error on the faint-end slope. Moreover, if we
assume a $5\%$ ($10\%$) error on the photometric redshift of all
objects (both faint and bright) we add $5\%$ ($10\%$) to the error on
the redshift evolution; $10\%$ ($20\%$) to the error on the faint-end
slope and $5\%$ ($10\%$) to the error on the bright-end slope.

It seems very likely that at least one ``dark energy'' mission will
fly in the next decade. We have shown here that several of the
proposed designs for such missions are also very well-suited to
addressing a completely different but important science goal:
constraining the formation of massive black holes and luminous QSOs at
the earliest epochs in the history of our Universe.

\section*{Acknowledgments}
We warmly thank Francesco Shankar for providing the SM07 estimate of
the high-z luminosity function. We also thank Bertrand Goldman and
Wolfgang Brandner for enlighting discussion on the brown dwarf
population. We are grateful to Hans-Walter Rix and Alejo
Mart{\'{\i}}nez-Sansigre for discussions and suggestions that helped
us to improve the paper. Some of the computations were carried out on
the PIA cluster of the Max-Planck-Institut f\"ur Astronomie at the
Rechenzentrum Garching.

\begin{table*}
\begin{center}
\begin{tabular}{cccccccccc}
\hline
& \multicolumn{3}{c}{Non-Evolving LF} & \multicolumn{3}{c}{Evolving LF}& \multicolumn{3}{c}{Edd. Accr. LF} \\
& F07a & F07b & SM07 &F07a & F07b & SM07 &F07a & F07b & SM07 \\
\hline
\multicolumn{10}{c}{$m_H < 23$} \\
\hline
$ 5.0<z< 5.5$ & $ 1.7               $ & $ 1.4               $ & $ 2.8               $ & $ 1.3               $ & $ 1.1               $ & $ 2.2               $ & $ 1.2               $ & $ 1.1               $ & $ 2.4               $ \\
$ 5.5<z< 6.0$ & $ 1.4               $ & $ 1.2               $ & $ 2.1               $ & $ 0.6               $ & $ 0.5               $ & $ 1.0               $ & $ 0.5               $ & $ 0.5               $ & $ 1.0               $ \\
$ 6.0<z< 6.5$ & $ 1.2               $ & $ 1.0               $ & $ 1.6               $ & $ 0.2               $ & $ 0.2               $ & $ 0.4               $ & $ 0.1               $ & $ 0.1               $ & $ 0.2               $ \\
$ 6.5<z< 7.0$ & $ 1.0               $ & $ 0.9               $ & $ 1.2               $ & $ 0.1               $ & $ 0.1               $ & $ 0.2               $ & $ 1.3 \times 10^{-2}$ & $ 1.6 \times 10^{-2}$ & $ 7.5 \times 10^{-3}$ \\
$ 7.0<z< 7.5$ & $ 0.9               $ & $ 0.8               $ & $ 1.0               $ & $ 4.9 \times 10^{-2}$ & $ 4.9 \times 10^{-2}$ & $ 8.4 \times 10^{-2}$ & $ 5.6 \times 10^{-4}$ & $ 6.5 \times 10^{-4}$ & $ 3.3 \times 10^{-4}$ \\
$ 7.5<z< 8.0$ & $ 0.8               $ & $ 0.7               $ & $ 0.8               $ & $ 2.2 \times 10^{-2}$ & $ 2.3 \times 10^{-2}$ & $ 3.8 \times 10^{-2}$ & --- & --- & --- \\
$ 8.0<z< 8.5$ & $ 0.7               $ & $ 0.6               $ & $ 0.7               $ & $ 1.1 \times 10^{-2}$ & $ 1.1 \times 10^{-2}$ & $ 2.0 \times 10^{-2}$ & --- & --- & --- \\
$ 8.5<z< 9.0$ & $ 0.6               $ & $ 0.5               $ & $ 0.5               $ & $ 4.7 \times 10^{-3}$ & $ 5.2 \times 10^{-3}$ & $ 8.6 \times 10^{-3}$ & --- & --- & --- \\
$ 9.0<z< 9.5$ & $ 0.5               $ & $ 0.5               $ & $ 0.4               $ & $ 2.2 \times 10^{-3}$ & $ 2.5 \times 10^{-3}$ & $ 4.0 \times 10^{-3}$ & --- & --- & --- \\
$ 9.5<z<10.0$ & $ 0.4               $ & $ 0.4               $ & $ 0.4               $ & $ 9.8 \times 10^{-4}$ & $ 1.2 \times 10^{-3}$ & $ 1.9 \times 10^{-3}$ & --- & --- & --- \\
$10.0<z<10.5$ & $ 0.4               $ & $ 0.4               $ & $ 0.3               $ & $ 4.6 \times 10^{-4}$ & $ 5.6 \times 10^{-4}$ & $ 9.8 \times 10^{-4}$ & --- & --- & --- \\
$10.5<z<11.0$ & $ 0.3               $ & $ 0.3               $ & $ 0.2               $ & $ 1.9 \times 10^{-4}$ & $ 2.5 \times 10^{-4}$ & $ 3.5 \times 10^{-4}$ & --- & --- & --- \\
$11.0<z<11.5$ & $ 0.2               $ & $ 0.2               $ & $ 0.1               $ & --- & --- & $ 1.2 \times 10^{-4}$ & --- & --- & --- \\
$11.5<z<12.0$ & $ 0.1               $ & $ 0.1               $ & $ 6.7 \times 10^{-2}$ & --- & --- & --- & --- & --- & --- \\
\hline
\multicolumn{10}{c}{$m_H < 26$} \\
\hline
$ 5.0<z< 5.5$ & $14.9               $ & $ 6.9               $ & $ 1.2 \times 10^{2} $ & $10.9               $ & $ 5.3               $ & $94.3               $ & $10.3               $ & $ 5.3               $ & $ 1.0 \times 10^{2} $ \\
$ 5.5<z< 6.0$ & $12.2               $ & $ 6.1               $ & $92.9               $ & $ 4.7               $ & $ 2.4               $ & $42.2               $ & $ 4.3               $ & $ 2.4               $ & $43.5               $ \\
$ 6.0<z< 6.5$ & $10.6               $ & $ 5.3               $ & $74.7               $ & $ 2.1               $ & $ 1.2               $ & $19.3               $ & $ 1.4               $ & $ 0.9               $ & $10.1               $ \\
$ 6.5<z< 7.0$ & $ 9.0               $ & $ 4.7               $ & $58.5               $ & $ 1.0               $ & $ 0.6               $ & $ 8.8               $ & $ 0.3               $ & $ 0.3               $ & $ 0.7               $ \\
$ 7.0<z< 7.5$ & $ 7.9               $ & $ 4.2               $ & $47.6               $ & $ 0.4               $ & $ 0.3               $ & $ 4.2               $ & $ 6.1 \times 10^{-2}$ & $ 7.1 \times 10^{-2}$ & $ 5.5 \times 10^{-2}$ \\
$ 7.5<z< 8.0$ & $ 7.0               $ & $ 3.8               $ & $39.8               $ & $ 0.2               $ & $ 0.1               $ & $ 2.0               $ & $ 8.7 \times 10^{-3}$ & $ 1.1 \times 10^{-2}$ & $ 4.7 \times 10^{-3}$ \\
$ 8.0<z< 8.5$ & $ 6.3               $ & $ 3.5               $ & $36.5               $ & $ 9.6 \times 10^{-2}$ & $ 6.3 \times 10^{-2}$ & $ 1.0               $ & $ 9.0 \times 10^{-4}$ & $ 1.0 \times 10^{-3}$ & $ 5.1 \times 10^{-4}$ \\
$ 8.5<z< 9.0$ & $ 5.5               $ & $ 3.1               $ & $28.3               $ & $ 4.4 \times 10^{-2}$ & $ 3.1 \times 10^{-2}$ & $ 0.5               $ & --- & --- & --- \\
$ 9.0<z< 9.5$ & $ 4.8               $ & $ 2.8               $ & $23.2               $ & $ 2.1 \times 10^{-2}$ & $ 1.5 \times 10^{-2}$ & $ 0.2               $ & --- & --- & --- \\
$ 9.5<z<10.0$ & $ 4.3               $ & $ 2.5               $ & $20.2               $ & $ 9.3 \times 10^{-3}$ & $ 7.3 \times 10^{-3}$ & $ 0.1               $ & --- & --- & --- \\
$10.0<z<10.5$ & $ 3.9               $ & $ 2.3               $ & $18.0               $ & $ 5.1 \times 10^{-3}$ & $ 3.6 \times 10^{-3}$ & $ 5.7 \times 10^{-2}$ & --- & --- & --- \\
$10.5<z<11.0$ & $ 3.1               $ & $ 1.9               $ & $12.0               $ & $ 2.0 \times 10^{-3}$ & $ 1.7 \times 10^{-3}$ & $ 2.2 \times 10^{-2}$ & --- & --- & --- \\
$11.0<z<11.5$ & $ 2.4               $ & $ 1.6               $ & $ 7.8               $ & $ 8.8 \times 10^{-4}$ & $ 7.7 \times 10^{-4}$ & $ 8.3 \times 10^{-3}$ & --- & --- & --- \\
$11.5<z<12.0$ & $ 1.8               $ & $ 1.2               $ & $ 5.2               $ & $ 3.3 \times 10^{-4}$ & $ 3.1 \times 10^{-4}$ & $ 3.2 \times 10^{-3}$ & --- & --- & --- \\
\hline
\multicolumn{10}{c}{$m_H < 29$} \\
\hline
$ 5.0<z< 5.5$ & $ 1.1 \times 10^{2} $ & $28.5               $ & $ 3.6 \times 10^{3} $ & $80.2               $ & $21.7               $ & $ 2.8 \times 10^{3} $ & $74.9               $ & $21.6               $ & $ 3.0 \times 10^{3} $ \\
$ 5.5<z< 6.0$ & $90.9               $ & $25.0               $ & $ 2.8 \times 10^{3} $ & $34.6               $ & $10.1               $ & $ 1.3 \times 10^{3} $ & $31.6               $ & $10.0               $ & $ 1.3 \times 10^{3} $ \\
$ 6.0<z< 6.5$ & $75.0               $ & $22.0               $ & $ 2.3 \times 10^{3} $ & $15.7               $ & $ 4.8               $ & $ 5.8 \times 10^{2} $ & $10.3               $ & $ 4.0               $ & $ 3.2 \times 10^{2} $ \\
$ 6.5<z< 7.0$ & $66.1               $ & $19.6               $ & $ 1.8 \times 10^{3} $ & $ 6.9               $ & $ 2.3               $ & $ 2.7 \times 10^{2} $ & $ 2.6               $ & $ 1.5               $ & $30.3               $ \\
$ 7.0<z< 7.5$ & $58.2               $ & $17.7               $ & $ 1.5 \times 10^{3} $ & $ 3.0               $ & $ 1.1               $ & $ 1.3 \times 10^{2} $ & $ 0.8               $ & $ 0.6               $ & $ 3.6               $ \\
$ 7.5<z< 8.0$ & $50.5               $ & $16.0               $ & $ 1.2 \times 10^{3} $ & $ 1.5               $ & $ 0.5               $ & $61.0               $ & $ 0.2               $ & $ 0.2               $ & $ 0.5               $ \\
$ 8.0<z< 8.5$ & $46.4               $ & $14.8               $ & $ 1.1 \times 10^{3} $ & $ 0.7               $ & $ 0.3               $ & $32.6               $ & $ 7.8 \times 10^{-2}$ & $ 8.9 \times 10^{-2}$ & $ 7.9 \times 10^{-2}$ \\
$ 8.5<z< 9.0$ & $40.1               $ & $13.1               $ & $ 9.0 \times 10^{2} $ & $ 0.3               $ & $ 0.1               $ & $14.7               $ & $ 1.9 \times 10^{-2}$ & $ 2.3 \times 10^{-2}$ & $ 1.2 \times 10^{-2}$ \\
$ 9.0<z< 9.5$ & $35.8               $ & $11.7               $ & $ 7.4 \times 10^{2} $ & $ 0.2               $ & $ 6.5 \times 10^{-2}$ & $ 7.1               $ & $ 3.7 \times 10^{-3}$ & $ 4.5 \times 10^{-3}$ & $ 1.9 \times 10^{-3}$ \\
$ 9.5<z<10.0$ & $32.5               $ & $10.7               $ & $ 6.5 \times 10^{2} $ & $ 7.3 \times 10^{-2}$ & $ 3.1 \times 10^{-2}$ & $ 3.5               $ & $ 6.4 \times 10^{-4}$ & $ 7.3 \times 10^{-4}$ & $ 3.7 \times 10^{-4}$ \\
$10.0<z<10.5$ & $28.4               $ & $ 9.8               $ & $ 5.8 \times 10^{2} $ & $ 3.7 \times 10^{-2}$ & $ 1.5 \times 10^{-2}$ & $ 1.8               $ & $ 1.2 \times 10^{-4}$ & $ 1.4 \times 10^{-4}$ & --- \\
$10.5<z<11.0$ & $23.7               $ & $ 8.4               $ & $ 4.0 \times 10^{2} $ & $ 1.5 \times 10^{-2}$ & $ 7.4 \times 10^{-3}$ & $ 0.7               $ & --- & --- & --- \\
$11.0<z<11.5$ & $18.2               $ & $ 6.9               $ & $ 2.7 \times 10^{2} $ & $ 6.4 \times 10^{-3}$ & $ 3.4 \times 10^{-3}$ & $ 0.3               $ & --- & --- & --- \\
$11.5<z<12.0$ & $12.1               $ & $ 4.9               $ & $ 1.9 \times 10^{2} $ & $ 2.6 \times 10^{-3}$ & $ 1.4 \times 10^{-3}$ & $ 0.1               $ & --- & --- & --- \\
\hline
\multicolumn{10}{c}{$m_H < 32$} \\
\hline
$ 5.0<z< 5.5$ & $ 7.5 \times 10^{2} $ & $ 1.1 \times 10^{2} $ & $ 9.9 \times 10^{4} $ & $ 5.6 \times 10^{2} $ & $85.2               $ & $ 7.8 \times 10^{4} $ & $ 5.3 \times 10^{2} $ & $84.7               $ & $ 8.3 \times 10^{4} $ \\
$ 5.5<z< 6.0$ & $ 6.3 \times 10^{2} $ & $98.3               $ & $ 7.8 \times 10^{4} $ & $ 2.5 \times 10^{2} $ & $39.6               $ & $ 3.5 \times 10^{4} $ & $ 2.3 \times 10^{2} $ & $39.2               $ & $ 3.6 \times 10^{4} $ \\
$ 6.0<z< 6.5$ & $ 5.4 \times 10^{2} $ & $86.5               $ & $ 6.3 \times 10^{4} $ & $ 1.1 \times 10^{2} $ & $19.0               $ & $ 1.6 \times 10^{4} $ & $74.1               $ & $16.0               $ & $ 8.9 \times 10^{3} $ \\
$ 6.5<z< 7.0$ & $ 4.7 \times 10^{2} $ & $77.1               $ & $ 5.0 \times 10^{4} $ & $49.3               $ & $ 9.1               $ & $ 7.5 \times 10^{3} $ & $18.7               $ & $ 6.1               $ & $ 9.0 \times 10^{2} $ \\
$ 7.0<z< 7.5$ & $ 4.0 \times 10^{2} $ & $69.7               $ & $ 4.1 \times 10^{4} $ & $23.1               $ & $ 4.4               $ & $ 3.6 \times 10^{3} $ & $ 5.8               $ & $ 2.6               $ & $ 1.2 \times 10^{2} $ \\
$ 7.5<z< 8.0$ & $ 3.7 \times 10^{2} $ & $63.1               $ & $ 3.5 \times 10^{4} $ & $10.6               $ & $ 2.1               $ & $ 1.7 \times 10^{3} $ & $ 2.1               $ & $ 1.3               $ & $22.3               $ \\
$ 8.0<z< 8.5$ & $ 3.4 \times 10^{2} $ & $58.3               $ & $ 3.2 \times 10^{4} $ & $ 5.3               $ & $ 1.1               $ & $ 9.1 \times 10^{2} $ & $ 0.9               $ & $ 0.6               $ & $ 4.8               $ \\
$ 8.5<z< 9.0$ & $ 2.9 \times 10^{2} $ & $51.7               $ & $ 2.5 \times 10^{4} $ & $ 2.4               $ & $ 0.5               $ & $ 4.1 \times 10^{2} $ & $ 0.4               $ & $ 0.3               $ & $ 1.0               $ \\
$ 9.0<z< 9.5$ & $ 2.6 \times 10^{2} $ & $46.5               $ & $ 2.1 \times 10^{4} $ & $ 1.1               $ & $ 0.3               $ & $ 2.0 \times 10^{2} $ & $ 0.2               $ & $ 0.2               $ & $ 0.2               $ \\
$ 9.5<z<10.0$ & $ 2.2 \times 10^{2} $ & $42.3               $ & $ 1.8 \times 10^{4} $ & $ 0.5               $ & $ 0.1               $ & $99.6               $ & $ 6.4 \times 10^{-2}$ & $ 7.4 \times 10^{-2}$ & $ 6.0 \times 10^{-2}$ \\
$10.0<z<10.5$ & $ 2.1 \times 10^{2} $ & $39.0               $ & $ 1.6 \times 10^{4} $ & $ 0.2               $ & $ 6.1 \times 10^{-2}$ & $52.1               $ & $ 2.6 \times 10^{-2}$ & $ 3.2 \times 10^{-2}$ & $ 1.7 \times 10^{-2}$ \\
$10.5<z<11.0$ & $ 1.7 \times 10^{2} $ & $33.4               $ & $ 1.1 \times 10^{4} $ & $ 0.1               $ & $ 2.9 \times 10^{-2}$ & $21.0               $ & $ 7.4 \times 10^{-3}$ & $ 9.0 \times 10^{-3}$ & $ 3.9 \times 10^{-3}$ \\
$11.0<z<11.5$ & $ 1.3 \times 10^{2} $ & $27.7               $ & $ 7.7 \times 10^{3} $ & $ 4.6 \times 10^{-2}$ & $ 1.4 \times 10^{-2}$ & $ 8.1               $ & $ 1.5 \times 10^{-3}$ & $ 1.7 \times 10^{-3}$ & $ 7.9 \times 10^{-4}$ \\
$11.5<z<12.0$ & $86.7               $ & $19.4               $ & $ 5.4 \times 10^{3} $ & $ 1.7 \times 10^{-2}$ & $ 5.8 \times 10^{-3}$ & $ 3.3               $ & $ 2.8 \times 10^{-4}$ & $ 3.2 \times 10^{-4}$ & $ 1.7 \times 10^{-4}$ \\
\hline
\end{tabular}
\caption{Predicted QSOs counts at different redshift (deg$^{-2}$)}\label{tab:counts1}
\end{center}
\end{table*}

\begin{table*}
\begin{center}
\begin{tabular}{cccccccccc}
\hline
& \multicolumn{3}{c}{Non-Evolving LF} & \multicolumn{3}{c}{Evolving LF}& \multicolumn{3}{c}{Edd. Accr. LF} \\
& F07a & F07b & SM07 &F07a & F07b & SM07 &F07a & F07b & SM07 \\
\hline
\multicolumn{10}{c}{$5.5<z<6.5$} \\
\hline
$m_H<23.0$ & $ 2.4               $ & $ 2.0               $ & $ 3.4               $ & $ 0.7               $ & $ 0.7               $ & $ 1.3               $ & $ 0.6               $ & $ 0.6               $ & $ 1.1               $ \\
$m_H<24.0$ & $ 5.2               $ & $ 3.7               $ & $13.3               $ & $ 1.6               $ & $ 1.2               $ & $ 4.9               $ & $ 1.3               $ & $ 1.1               $ & $ 4.2               $ \\
$m_H<25.0$ & $10.7               $ & $ 6.4               $ & $46.9               $ & $ 3.2               $ & $ 2.0               $ & $17.3               $ & $ 2.7               $ & $ 1.9               $ & $14.9               $ \\
$m_H<26.0$ & $21.3               $ & $10.6               $ & $ 1.5 \times 10^{2} $ & $ 6.3               $ & $ 3.4               $ & $57.2               $ & $ 5.3               $ & $ 3.1               $ & $49.2               $ \\
$m_H<27.0$ & $39.9               $ & $16.8               $ & $ 4.9 \times 10^{2} $ & $12.5               $ & $ 5.5               $ & $ 1.8 \times 10^{2} $ & $10.3               $ & $ 5.0               $ & $ 1.6 \times 10^{2} $ \\
$m_H<28.0$ & $78.6               $ & $27.2               $ & $ 1.5 \times 10^{3} $ & $23.8               $ & $ 8.7               $ & $ 5.6 \times 10^{2} $ & $20.0               $ & $ 8.1               $ & $ 4.8 \times 10^{2} $ \\
$m_H<29.0$ & $ 1.5 \times 10^{2} $ & $42.8               $ & $ 4.7 \times 10^{3} $ & $45.2               $ & $13.7               $ & $ 1.7 \times 10^{3} $ & $38.7               $ & $12.9               $ & $ 1.5 \times 10^{3} $ \\
$m_H<30.0$ & $ 2.9 \times 10^{2} $ & $68.2               $ & $ 1.4 \times 10^{4} $ & $89.2               $ & $22.0               $ & $ 5.2 \times 10^{3} $ & $74.6               $ & $20.5               $ & $ 4.5 \times 10^{3} $ \\
$m_H<31.0$ & $ 5.7 \times 10^{2} $ & $ 1.1 \times 10^{2} $ & $ 4.3 \times 10^{4} $ & $ 1.7 \times 10^{2} $ & $34.8               $ & $ 1.6 \times 10^{4} $ & $ 1.4 \times 10^{2} $ & $32.3               $ & $ 1.4 \times 10^{4} $ \\
$m_H<32.0$ & $ 1.1 \times 10^{3} $ & $ 1.7 \times 10^{2} $ & $ 1.3 \times 10^{5} $ & $ 3.3 \times 10^{2} $ & $54.7               $ & $ 4.8 \times 10^{4} $ & $ 2.8 \times 10^{2} $ & $50.9               $ & $ 4.1 \times 10^{4} $ \\
\hline
\multicolumn{10}{c}{$6.5<z<7.5$} \\
\hline
$m_H<23.0$ & $ 1.8               $ & $ 1.5               $ & $ 2.0               $ & $ 0.1               $ & $ 0.1               $ & $ 0.2               $ & $ 1.3 \times 10^{-2}$ & $ 1.6 \times 10^{-2}$ & $ 7.7 \times 10^{-3}$ \\
$m_H<24.0$ & $ 3.9               $ & $ 2.8               $ & $ 8.0               $ & $ 0.3               $ & $ 0.3               $ & $ 1.0               $ & $ 4.9 \times 10^{-2}$ & $ 5.8 \times 10^{-2}$ & $ 4.0 \times 10^{-2}$ \\
$m_H<25.0$ & $ 7.7               $ & $ 5.0               $ & $28.8               $ & $ 0.6               $ & $ 0.5               $ & $ 3.6               $ & $ 0.1               $ & $ 0.2               $ & $ 0.2               $ \\
$m_H<26.0$ & $15.2               $ & $ 8.3               $ & $97.0               $ & $ 1.3               $ & $ 0.8               $ & $11.9               $ & $ 0.3               $ & $ 0.3               $ & $ 0.8               $ \\
$m_H<27.0$ & $30.1               $ & $13.2               $ & $ 3.1 \times 10^{2} $ & $ 2.5               $ & $ 1.2               $ & $38.3               $ & $ 0.8               $ & $ 0.6               $ & $ 2.9               $ \\
$m_H<28.0$ & $59.5               $ & $21.4               $ & $ 9.7 \times 10^{2} $ & $ 4.9               $ & $ 2.0               $ & $ 1.2 \times 10^{2} $ & $ 1.6               $ & $ 1.1               $ & $ 9.9               $ \\
$m_H<29.0$ & $ 1.1 \times 10^{2} $ & $33.8               $ & $ 3.0 \times 10^{3} $ & $ 9.2               $ & $ 3.1               $ & $ 3.7 \times 10^{2} $ & $ 3.1               $ & $ 1.9               $ & $32.6               $ \\
$m_H<30.0$ & $ 2.2 \times 10^{2} $ & $53.8               $ & $ 9.1 \times 10^{3} $ & $18.3               $ & $ 5.0               $ & $ 1.1 \times 10^{3} $ & $ 6.1               $ & $ 3.1               $ & $ 1.0 \times 10^{2} $ \\
$m_H<31.0$ & $ 4.2 \times 10^{2} $ & $85.9               $ & $ 2.8 \times 10^{4} $ & $36.8               $ & $ 8.0               $ & $ 3.4 \times 10^{3} $ & $11.9               $ & $ 5.1               $ & $ 3.2 \times 10^{2} $ \\
$m_H<32.0$ & $ 8.1 \times 10^{2} $ & $ 1.3 \times 10^{2} $ & $ 8.3 \times 10^{4} $ & $69.1               $ & $12.5               $ & $ 1.0 \times 10^{4} $ & $23.0               $ & $ 8.1               $ & $ 9.8 \times 10^{2} $ \\
\hline
\multicolumn{10}{c}{$7.5<z<8.5$} \\
\hline
$m_H<23.0$ & $ 1.3               $ & $ 1.2               $ & $ 1.4               $ & $ 3.0 \times 10^{-2}$ & $ 3.2 \times 10^{-2}$ & $ 5.4 \times 10^{-2}$ & --- & --- & --- \\
$m_H<24.0$ & $ 2.9               $ & $ 2.3               $ & $ 5.6               $ & $ 6.6 \times 10^{-2}$ & $ 6.0 \times 10^{-2}$ & $ 0.2               $ & $ 2.0 \times 10^{-4}$ & $ 2.3 \times 10^{-4}$ & $ 1.3 \times 10^{-4}$ \\
$m_H<25.0$ & $ 5.9               $ & $ 4.0               $ & $20.5               $ & $ 0.1               $ & $ 0.1               $ & $ 0.8               $ & $ 1.6 \times 10^{-3}$ & $ 1.8 \times 10^{-3}$ & $ 8.6 \times 10^{-4}$ \\
$m_H<26.0$ & $12.0               $ & $ 6.8               $ & $69.8               $ & $ 0.3               $ & $ 0.2               $ & $ 2.8               $ & $ 9.2 \times 10^{-3}$ & $ 1.1 \times 10^{-2}$ & $ 5.0 \times 10^{-3}$ \\
$m_H<27.0$ & $23.5               $ & $10.8               $ & $ 2.3 \times 10^{2} $ & $ 0.5               $ & $ 0.3               $ & $ 9.0               $ & $ 3.9 \times 10^{-2}$ & $ 4.7 \times 10^{-2}$ & $ 2.7 \times 10^{-2}$ \\
$m_H<28.0$ & $45.1               $ & $17.6               $ & $ 7.1 \times 10^{2} $ & $ 1.0               $ & $ 0.5               $ & $28.0               $ & $ 0.1               $ & $ 0.1               $ & $ 0.1               $ \\
$m_H<29.0$ & $88.5               $ & $27.8               $ & $ 2.2 \times 10^{3} $ & $ 2.0               $ & $ 0.7               $ & $86.3               $ & $ 0.3               $ & $ 0.3               $ & $ 0.6               $ \\
$m_H<30.0$ & $ 1.7 \times 10^{2} $ & $44.3               $ & $ 6.6 \times 10^{3} $ & $ 3.7               $ & $ 1.2               $ & $ 2.6 \times 10^{2} $ & $ 0.7               $ & $ 0.6               $ & $ 2.2               $ \\
$m_H<31.0$ & $ 3.4 \times 10^{2} $ & $70.8               $ & $ 2.0 \times 10^{4} $ & $ 7.3               $ & $ 1.9               $ & $ 8.0 \times 10^{2} $ & $ 1.4               $ & $ 1.0               $ & $ 7.7               $ \\
$m_H<32.0$ & $ 6.3 \times 10^{2} $ & $ 1.1 \times 10^{2} $ & $ 6.1 \times 10^{4} $ & $14.1               $ & $ 3.0               $ & $ 2.4 \times 10^{3} $ & $ 2.8               $ & $ 1.7               $ & $25.6               $ \\
\hline
\multicolumn{10}{c}{$8.5<z<9.5$} \\
\hline
$m_H<23.0$ & $ 0.9               $ & $ 0.9               $ & $ 0.9               $ & $ 6.2 \times 10^{-3}$ & $ 7.1 \times 10^{-3}$ & $ 1.2 \times 10^{-2}$ & --- & --- & --- \\
$m_H<24.0$ & $ 2.2               $ & $ 1.8               $ & $ 3.6               $ & $ 1.4 \times 10^{-2}$ & $ 1.4 \times 10^{-2}$ & $ 4.9 \times 10^{-2}$ & --- & --- & --- \\
$m_H<25.0$ & $ 4.7               $ & $ 3.2               $ & $13.6               $ & $ 3.0 \times 10^{-2}$ & $ 2.5 \times 10^{-2}$ & $ 0.2               $ & --- & --- & --- \\
$m_H<26.0$ & $ 9.5               $ & $ 5.4               $ & $46.9               $ & $ 5.9 \times 10^{-2}$ & $ 4.2 \times 10^{-2}$ & $ 0.6               $ & --- & --- & --- \\
$m_H<27.0$ & $18.0               $ & $ 8.7               $ & $ 1.5 \times 10^{2} $ & $ 0.1               $ & $ 7.0 \times 10^{-2}$ & $ 2.1               $ & $ 6.4 \times 10^{-4}$ & $ 7.4 \times 10^{-4}$ & $ 3.8 \times 10^{-4}$ \\
$m_H<28.0$ & $35.6               $ & $14.2               $ & $ 4.8 \times 10^{2} $ & $ 0.2               $ & $ 0.1               $ & $ 6.5               $ & $ 4.4 \times 10^{-3}$ & $ 5.2 \times 10^{-3}$ & $ 2.3 \times 10^{-3}$ \\
$m_H<29.0$ & $67.7               $ & $22.5               $ & $ 1.5 \times 10^{3} $ & $ 0.4               $ & $ 0.2               $ & $20.0               $ & $ 2.2 \times 10^{-2}$ & $ 2.6 \times 10^{-2}$ & $ 1.3 \times 10^{-2}$ \\
$m_H<30.0$ & $ 1.3 \times 10^{2} $ & $35.9               $ & $ 4.6 \times 10^{3} $ & $ 0.9               $ & $ 0.3               $ & $61.1               $ & $ 7.6 \times 10^{-2}$ & $ 8.9 \times 10^{-2}$ & $ 6.5 \times 10^{-2}$ \\
$m_H<31.0$ & $ 2.6 \times 10^{2} $ & $57.4               $ & $ 1.4 \times 10^{4} $ & $ 1.7               $ & $ 0.5               $ & $ 1.9 \times 10^{2} $ & $ 0.2               $ & $ 0.2               $ & $ 0.3               $ \\
$m_H<32.0$ & $ 4.9 \times 10^{2} $ & $90.0               $ & $ 4.2 \times 10^{4} $ & $ 3.2               $ & $ 0.7               $ & $ 5.6 \times 10^{2} $ & $ 0.5               $ & $ 0.4               $ & $ 1.2               $ \\
\hline
\end{tabular}
\caption{Predicted QSOs counts at different $H$-band limits (deg$^{-2}$)}\label{tab:counts2}
\end{center}
\end{table*}

\begin{table*}
\begin{center}
\begin{tabular}{cccccccccc}
\hline
& \multicolumn{3}{c}{Non-Evolving LF} & \multicolumn{3}{c}{Evolving LF}& \multicolumn{3}{c}{Edd. Accr. LF} \\
& F07a & F07b & SM07 &F07a & F07b & SM07 &F07a & F07b & SM07 \\
\hline
\multicolumn{10}{c}{$S(2-10 KeV) < 10^{-15.5} erg/s/cm^{-2}$} \\
\hline
$ 5.0<z< 5.5$ & $ 3.0               $ & $ 1.7               $ & $ 6.6               $ & $ 2.1               $ & $ 1.2               $ & $ 5.3               $ & $ 2.0               $ & $ 1.6               $ & $ 5.8               $ \\
$ 5.5<z< 6.0$ & $ 2.3               $ & $ 1.3               $ & $ 4.0               $ & $ 0.9               $ & $ 0.5               $ & $ 1.8               $ & $ 0.8               $ & $ 0.7               $ & $ 2.1               $ \\
$ 6.0<z< 6.5$ & $ 1.7               $ & $ 1.0               $ & $ 2.5               $ & $ 0.3               $ & $ 0.2               $ & $ 0.7               $ & $ 0.2               $ & $ 0.2               $ & $ 0.4               $ \\
$ 6.5<z< 7.0$ & $ 1.2               $ & $ 0.7               $ & $ 1.6               $ & $ 0.1               $ & $ 8.8 \times 10^{-2}$ & $ 0.2               $ & $ 1.8 \times 10^{-2}$ & $ 2.2 \times 10^{-2}$ & $ 1.1 \times 10^{-2}$ \\
$ 7.0<z< 7.5$ & $ 1.0               $ & $ 0.6               $ & $ 1.0               $ & $ 5.7 \times 10^{-2}$ & $ 3.9 \times 10^{-2}$ & $ 8.8 \times 10^{-2}$ & $ 7.2 \times 10^{-4}$ & $ 8.3 \times 10^{-4}$ & $ 4.1 \times 10^{-4}$ \\
$ 7.5<z< 8.0$ & $ 0.7               $ & $ 0.5               $ & $ 0.7               $ & $ 2.2 \times 10^{-2}$ & $ 1.6 \times 10^{-2}$ & $ 3.4 \times 10^{-2}$ & ---     & ---     & ---     \\
$ 8.0<z< 8.5$ & $ 0.6               $ & $ 0.4               $ & $ 0.5               $ & $ 9.7 \times 10^{-3}$ & $ 7.1 \times 10^{-3}$ & $ 1.4 \times 10^{-2}$ & ---     & ---     & ---     \\
$ 8.5<z< 9.0$ & $ 0.5               $ & $ 0.3               $ & $ 0.3               $ & $ 4.1 \times 10^{-3}$ & $ 3.0 \times 10^{-3}$ & $ 5.4 \times 10^{-3}$ & ---     & ---     & ---     \\
$ 9.0<z< 9.5$ & $ 0.4               $ & $ 0.2               $ & $ 0.2               $ & $ 1.5 \times 10^{-3}$ & $ 1.2 \times 10^{-3}$ & $ 2.2 \times 10^{-3}$ & ---     & ---     & ---     \\
$ 9.5<z<10.0$ & $ 0.3               $ & $ 0.2               $ & $ 0.2               $ & $ 7.7 \times 10^{-4}$ & $ 5.6 \times 10^{-4}$ & $ 9.2 \times 10^{-4}$ & ---     & ---     & ---     \\
$10.0<z<10.5$ & $ 0.2               $ & $ 0.1               $ & $ 0.1               $ & $ 3.2 \times 10^{-4}$ & $ 2.3 \times 10^{-4}$ & $ 3.7 \times 10^{-4}$ & ---     & ---     & ---     \\
$10.5<z<11.0$ & $ 0.2               $ & $ 0.1               $ & $ 8.4 \times 10^{-2}$ & $ 1.2 \times 10^{-4}$ & ---     & $ 1.5 \times 10^{-4}$ & ---     & ---     & ---     \\
$11.0<z<11.5$ & $ 0.1               $ & $ 8.2 \times 10^{-2}$ & $ 6.1 \times 10^{-2}$ & ---     & ---     & ---     & ---     & ---     & ---     \\
$11.5<z<12.0$ & $ 0.1               $ & $ 6.1 \times 10^{-2}$ & $ 4.4 \times 10^{-2}$ & ---     & ---     & ---     & ---     & ---     & ---     \\
\hline
\multicolumn{10}{c}{$S(2-10 KeV) < 10^{-16.5} erg/s/cm^{-2}$} \\
\hline
$ 5.0<z< 5.5$ & $27.3               $ & $11.2               $ & $ 3.3 \times 10^{2} $ & $20.8               $ & $ 8.6               $ & $ 2.6 \times 10^{2} $ & $18.6               $ & $ 8.1               $ & $ 2.9 \times 10^{2} $ \\
$ 5.5<z< 6.0$ & $22.3               $ & $ 9.5               $ & $ 2.1 \times 10^{2} $ & $ 8.0               $ & $ 3.7               $ & $96.7               $ & $ 7.2               $ & $ 3.5               $ & $ 1.1 \times 10^{2} $ \\
$ 6.0<z< 6.5$ & $16.7               $ & $ 7.7               $ & $ 1.4 \times 10^{2} $ & $ 3.8               $ & $ 1.8               $ & $36.7               $ & $ 2.1               $ & $ 1.3               $ & $22.8               $ \\
$ 6.5<z< 7.0$ & $13.6               $ & $ 6.4               $ & $ 1.0 \times 10^{2} $ & $ 1.2               $ & $ 0.7               $ & $15.4               $ & $ 0.4               $ & $ 0.4               $ & $ 1.4               $ \\
$ 7.0<z< 7.5$ & $11.2               $ & $ 5.5               $ & $73.2               $ & $ 0.6               $ & $ 0.3               $ & $ 6.4               $ & $ 9.0 \times 10^{-2}$ & $ 0.1               $ & $10.0 \times 10^{-2}$ \\
$ 7.5<z< 8.0$ & $ 9.1               $ & $ 4.7               $ & $51.0               $ & $ 0.3               $ & $ 0.2               $ & $ 2.6               $ & $ 1.3 \times 10^{-2}$ & $ 1.6 \times 10^{-2}$ & $ 7.2 \times 10^{-3}$ \\
$ 8.0<z< 8.5$ & $ 7.2               $ & $ 3.9               $ & $36.5               $ & $ 9.8 \times 10^{-2}$ & $ 6.8 \times 10^{-2}$ & $ 1.0               $ & $ 1.1 \times 10^{-3}$ & $ 1.3 \times 10^{-3}$ & $ 6.0 \times 10^{-4}$ \\
$ 8.5<z< 9.0$ & $ 5.7               $ & $ 3.3               $ & $27.3               $ & $ 5.2 \times 10^{-2}$ & $ 3.5 \times 10^{-2}$ & $ 0.5               $ & ---     & ---     & ---     \\
$ 9.0<z< 9.5$ & $ 5.0               $ & $ 2.9               $ & $21.2               $ & $ 2.4 \times 10^{-2}$ & $ 1.7 \times 10^{-2}$ & $ 0.2               $ & ---     & ---     & ---     \\
$ 9.5<z<10.0$ & $ 3.9               $ & $ 2.4               $ & $16.6               $ & $ 1.0 \times 10^{-2}$ & $ 7.5 \times 10^{-3}$ & $ 9.1 \times 10^{-2}$ & ---     & ---     & ---     \\
$10.0<z<10.5$ & $ 3.6               $ & $ 2.2               $ & $12.4               $ & $ 4.0 \times 10^{-3}$ & $ 3.4 \times 10^{-3}$ & $ 3.9 \times 10^{-2}$ & ---     & ---     & ---     \\
$10.5<z<11.0$ & $ 2.8               $ & $ 1.8               $ & $ 9.7               $ & $ 1.8 \times 10^{-3}$ & $ 1.5 \times 10^{-3}$ & $ 1.8 \times 10^{-2}$ & ---     & ---     & ---     \\
$11.0<z<11.5$ & $ 2.7               $ & $ 1.7               $ & $ 7.5               $ & $ 9.6 \times 10^{-4}$ & $ 8.1 \times 10^{-4}$ & $ 8.0 \times 10^{-3}$ & ---     & ---     & ---     \\
$11.5<z<12.0$ & $ 2.2               $ & $ 1.5               $ & $ 6.1               $ & $ 4.4 \times 10^{-4}$ & $ 3.9 \times 10^{-4}$ & $ 3.7 \times 10^{-3}$ & ---     & ---     & ---     \\
\hline
\multicolumn{10}{c}{$S(2-10 KeV) < 10^{-17.5} erg/s/cm^{-2}$} \\
\hline
$ 5.0<z< 5.5$ & $ 2.2 \times 10^{2} $ & $48.8               $ & $ 1.0 \times 10^{4} $ & $ 1.6 \times 10^{2} $ & $35.9               $ & $ 8.0 \times 10^{3} $ & $ 1.4 \times 10^{2} $ & $33.2               $ & $ 8.4 \times 10^{3} $ \\
$ 5.5<z< 6.0$ & $ 1.6 \times 10^{2} $ & $38.1               $ & $ 6.6 \times 10^{3} $ & $62.3               $ & $15.7               $ & $ 3.0 \times 10^{3} $ & $54.4               $ & $14.6               $ & $ 3.3 \times 10^{3} $ \\
$ 6.0<z< 6.5$ & $ 1.2 \times 10^{2} $ & $30.7               $ & $ 4.8 \times 10^{3} $ & $26.4               $ & $ 7.2               $ & $ 1.3 \times 10^{3} $ & $16.5               $ & $ 5.6               $ & $ 7.3 \times 10^{2} $ \\
$ 6.5<z< 7.0$ & $ 1.1 \times 10^{2} $ & $27.8               $ & $ 3.4 \times 10^{3} $ & $11.8               $ & $ 3.4               $ & $ 5.1 \times 10^{2} $ & $ 3.8               $ & $ 2.0               $ & $60.9               $ \\
$ 7.0<z< 7.5$ & $88.9               $ & $24.2               $ & $ 2.4 \times 10^{3} $ & $ 4.3               $ & $ 1.4               $ & $ 2.1 \times 10^{2} $ & $ 1.1               $ & $ 0.8               $ & $ 6.7               $ \\
$ 7.5<z< 8.0$ & $72.0               $ & $20.5               $ & $ 1.8 \times 10^{3} $ & $ 1.8               $ & $ 0.6               $ & $90.5               $ & $ 0.3               $ & $ 0.3               $ & $ 0.8               $ \\
$ 8.0<z< 8.5$ & $59.9               $ & $17.7               $ & $ 1.4 \times 10^{3} $ & $ 0.9               $ & $ 0.3               $ & $39.7               $ & $ 9.7 \times 10^{-2}$ & $ 0.1               $ & $ 0.1               $ \\
$ 8.5<z< 9.0$ & $48.0               $ & $15.1               $ & $ 1.0 \times 10^{3} $ & $ 0.4               $ & $ 0.1               $ & $17.4               $ & $ 2.4 \times 10^{-2}$ & $ 2.9 \times 10^{-2}$ & $ 1.5 \times 10^{-2}$ \\
$ 9.0<z< 9.5$ & $39.5               $ & $13.0               $ & $ 8.3 \times 10^{2} $ & $ 0.2               $ & $ 7.7 \times 10^{-2}$ & $ 7.9               $ & $ 4.6 \times 10^{-3}$ & $ 5.5 \times 10^{-3}$ & $ 2.4 \times 10^{-3}$ \\
$ 9.5<z<10.0$ & $35.8               $ & $11.9               $ & $ 6.5 \times 10^{2} $ & $ 8.5 \times 10^{-2}$ & $ 3.5 \times 10^{-2}$ & $ 3.5               $ & $ 7.0 \times 10^{-4}$ & $ 8.1 \times 10^{-4}$ & $ 4.0 \times 10^{-4}$ \\
$10.0<z<10.5$ & $30.1               $ & $10.4               $ & $ 5.2 \times 10^{2} $ & $ 3.3 \times 10^{-2}$ & $ 1.6 \times 10^{-2}$ & $ 1.6               $ & $ 1.1 \times 10^{-4}$ & $ 1.2 \times 10^{-4}$ & ---     \\
$10.5<z<11.0$ & $25.8               $ & $ 9.2               $ & $ 4.1 \times 10^{2} $ & $ 1.5 \times 10^{-2}$ & $ 7.6 \times 10^{-3}$ & $ 0.8               $ & ---     & ---     & ---     \\
$11.0<z<11.5$ & $21.9               $ & $ 8.1               $ & $ 3.3 \times 10^{2} $ & $ 7.8 \times 10^{-3}$ & $ 3.8 \times 10^{-3}$ & $ 0.3               $ & ---     & ---     & ---     \\
$11.5<z<12.0$ & $19.7               $ & $ 7.4               $ & $ 2.7 \times 10^{2} $ & $ 3.5 \times 10^{-3}$ & $ 1.8 \times 10^{-3}$ & $ 0.2               $ & ---     & ---     & ---     \\
\hline
\multicolumn{10}{c}{$S(2-10 KeV) < 10^{-18.5} erg/s/cm^{-2}$} \\
\hline
$ 5.0<z< 5.5$ & $ 1.4 \times 10^{3} $ & $ 1.7 \times 10^{2} $ & $ 2.4 \times 10^{5} $ & $ 1.1 \times 10^{3} $ & $ 1.4 \times 10^{2} $ & $ 1.9 \times 10^{5} $ & $ 9.0 \times 10^{2} $ & $ 1.2 \times 10^{2} $ & $ 2.0 \times 10^{5} $ \\
$ 5.5<z< 6.0$ & $ 1.1 \times 10^{3} $ & $ 1.4 \times 10^{2} $ & $ 1.6 \times 10^{5} $ & $ 4.3 \times 10^{2} $ & $59.8               $ & $ 7.5 \times 10^{4} $ & $ 3.6 \times 10^{2} $ & $54.1               $ & $ 8.2 \times 10^{4} $ \\
$ 6.0<z< 6.5$ & $ 7.9 \times 10^{2} $ & $ 1.2 \times 10^{2} $ & $ 1.2 \times 10^{5} $ & $ 1.9 \times 10^{2} $ & $28.2               $ & $ 3.0 \times 10^{4} $ & $ 1.1 \times 10^{2} $ & $21.1               $ & $ 1.9 \times 10^{4} $ \\
$ 6.5<z< 7.0$ & $ 6.7 \times 10^{2} $ & $ 1.0 \times 10^{2} $ & $ 8.4 \times 10^{4} $ & $68.9               $ & $11.7               $ & $ 1.3 \times 10^{4} $ & $26.0               $ & $ 7.6               $ & $ 1.6 \times 10^{3} $ \\
$ 7.0<z< 7.5$ & $ 5.6 \times 10^{2} $ & $87.9               $ & $ 6.4 \times 10^{4} $ & $31.8               $ & $ 5.7               $ & $ 5.6 \times 10^{3} $ & $ 7.6               $ & $ 3.2               $ & $ 2.0 \times 10^{2} $ \\
$ 7.5<z< 8.0$ & $ 4.7 \times 10^{2} $ & $76.4               $ & $ 4.9 \times 10^{4} $ & $15.0               $ & $ 2.8               $ & $ 2.4 \times 10^{3} $ & $ 2.6               $ & $ 1.5               $ & $32.4               $ \\
$ 8.0<z< 8.5$ & $ 4.1 \times 10^{2} $ & $68.3               $ & $ 3.7 \times 10^{4} $ & $ 6.2               $ & $ 1.3               $ & $ 1.1 \times 10^{3} $ & $ 1.0               $ & $ 0.7               $ & $ 6.1               $ \\
$ 8.5<z< 9.0$ & $ 3.4 \times 10^{2} $ & $58.8               $ & $ 2.9 \times 10^{4} $ & $ 2.8               $ & $ 0.6               $ & $ 4.9 \times 10^{2} $ & $ 0.4               $ & $ 0.3               $ & $ 1.3               $ \\
$ 9.0<z< 9.5$ & $ 2.8 \times 10^{2} $ & $51.6               $ & $ 2.2 \times 10^{4} $ & $ 1.3               $ & $ 0.3               $ & $ 2.1 \times 10^{2} $ & $ 0.2               $ & $ 0.2               $ & $ 0.3               $ \\
$ 9.5<z<10.0$ & $ 2.6 \times 10^{2} $ & $47.0               $ & $ 1.8 \times 10^{4} $ & $ 0.6               $ & $ 0.1               $ & $99.6               $ & $ 6.6 \times 10^{-2}$ & $ 7.6 \times 10^{-2}$ & $ 6.4 \times 10^{-2}$ \\
$10.0<z<10.5$ & $ 2.2 \times 10^{2} $ & $42.0               $ & $ 1.5 \times 10^{4} $ & $ 0.3               $ & $ 6.6 \times 10^{-2}$ & $46.4               $ & $ 2.4 \times 10^{-2}$ & $ 2.9 \times 10^{-2}$ & $ 1.6 \times 10^{-2}$ \\
$10.5<z<11.0$ & $ 1.9 \times 10^{2} $ & $37.2               $ & $ 1.2 \times 10^{4} $ & $ 0.1               $ & $ 3.1 \times 10^{-2}$ & $21.4               $ & $ 7.7 \times 10^{-3}$ & $ 9.4 \times 10^{-3}$ & $ 4.1 \times 10^{-3}$ \\
$11.0<z<11.5$ & $ 1.6 \times 10^{2} $ & $32.7               $ & $ 9.7 \times 10^{3} $ & $ 6.5 \times 10^{-2}$ & $ 1.6 \times 10^{-2}$ & $10.2               $ & $ 2.2 \times 10^{-3}$ & $ 2.6 \times 10^{-3}$ & $ 1.1 \times 10^{-3}$ \\
$11.5<z<12.0$ & $ 1.3 \times 10^{2} $ & $29.1               $ & $ 8.0 \times 10^{3} $ & $ 2.6 \times 10^{-2}$ & $ 7.5 \times 10^{-3}$ & $ 4.7               $ & $ 6.0 \times 10^{-4}$ & $ 7.0 \times 10^{-4}$ & $ 3.5 \times 10^{-4}$ \\
\hline
\end{tabular}
\caption{Predicted QSOs counts at different redshift (deg$^{-2}$)}\label{tab:counts3}
\end{center}
\end{table*}

\begin{table*}
\begin{center}
\begin{tabular}{cccccccccc}
\hline
& \multicolumn{3}{c}{Non-Evolving LF} & \multicolumn{3}{c}{Evolving LF}& \multicolumn{3}{c}{Edd. Accr. LF} \\
& F07a & F07b & SM07 &F07a & F07b & SM07 &F07a & F07b & SM07 \\
\hline
\multicolumn{10}{c}{$5.5<z<6.5$} \\
\hline
$log(S)<-14.0$ & $ 4.2 \times 10^{-3}$ & $ 3.9 \times 10^{-3}$ & $ 1.4 \times 10^{-3}$ & $ 1.3 \times 10^{-3}$ & $ 1.3 \times 10^{-3}$ & $ 5.6 \times 10^{-4}$ & $ 1.1 \times 10^{-3}$ & $ 1.2 \times 10^{-3}$ & $ 6.1 \times 10^{-4}$ \\
$log(S)<-14.5$ & $ 9.3 \times 10^{-2}$ & $ 9.9 \times 10^{-2}$ & $ 3.2 \times 10^{-2}$ & $ 3.0 \times 10^{-2}$ & $ 3.4 \times 10^{-2}$ & $ 1.3 \times 10^{-2}$ & $ 2.6 \times 10^{-2}$ & $ 3.0 \times 10^{-2}$ & $ 1.3 \times 10^{-2}$ \\
$log(S)<-15.0$ & $ 0.8               $ & $ 0.8               $ & $ 0.5               $ & $ 0.3               $ & $ 0.3               $ & $ 0.2               $ & $ 0.2               $ & $ 0.2               $ & $ 0.2               $ \\
$log(S)<-15.5$ & $ 3.7               $ & $ 2.9               $ & $ 6.0               $ & $ 1.1               $ & $ 0.9               $ & $ 2.3               $ & $ 0.9               $ & $ 0.8               $ & $ 2.2               $ \\
$log(S)<-16.0$ & $12.4               $ & $ 7.2               $ & $49.2               $ & $ 4.0               $ & $ 2.4               $ & $19.0               $ & $ 2.9               $ & $ 2.0               $ & $17.9               $ \\
$log(S)<-16.5$ & $36.3               $ & $15.8               $ & $ 3.3 \times 10^{2} $ & $11.3               $ & $ 5.2               $ & $ 1.2 \times 10^{2} $ & $ 8.6               $ & $ 4.4               $ & $ 1.2 \times 10^{2} $ \\
$log(S)<-17.0$ & $ 1.0 \times 10^{2} $ & $33.2               $ & $ 1.9 \times 10^{3} $ & $29.3               $ & $10.3               $ & $ 7.3 \times 10^{2} $ & $24.2               $ & $ 9.2               $ & $ 6.9 \times 10^{2} $ \\
$log(S)<-17.5$ & $ 2.7 \times 10^{2} $ & $65.4               $ & $ 1.0 \times 10^{4} $ & $80.3               $ & $21.0               $ & $ 3.9 \times 10^{3} $ & $65.5               $ & $18.6               $ & $ 3.8 \times 10^{3} $ \\
$log(S)<-18.0$ & $ 6.8 \times 10^{2} $ & $ 1.3 \times 10^{2} $ & $ 5.4 \times 10^{4} $ & $ 2.2 \times 10^{2} $ & $41.4               $ & $ 2.1 \times 10^{4} $ & $ 1.7 \times 10^{2} $ & $36.4               $ & $ 1.9 \times 10^{4} $ \\
$log(S)<-18.5$ & $ 1.8 \times 10^{3} $ & $ 2.5 \times 10^{2} $ & $ 2.6 \times 10^{5} $ & $ 5.4 \times 10^{2} $ & $78.7               $ & $ 9.8 \times 10^{4} $ & $ 4.3 \times 10^{2} $ & $69.4               $ & $ 9.2 \times 10^{4} $ \\
\hline
\multicolumn{10}{c}{$6.5<z<7.5$} \\
\hline
$log(S)<-14.0$ & $ 1.1 \times 10^{-3}$ & $ 1.0 \times 10^{-3}$ & $ 4.1 \times 10^{-4}$ & ---     & $ 1.0 \times 10^{-4}$ & ---     & ---     & ---     & ---     \\
$log(S)<-14.5$ & $ 3.0 \times 10^{-2}$ & $ 3.1 \times 10^{-2}$ & $ 1.0 \times 10^{-2}$ & $ 2.9 \times 10^{-3}$ & $ 3.3 \times 10^{-3}$ & $ 1.3 \times 10^{-3}$ & ---     & ---     & ---     \\
$log(S)<-15.0$ & $ 0.4               $ & $ 0.4               $ & $ 0.2               $ & $ 3.5 \times 10^{-2}$ & $ 4.1 \times 10^{-2}$ & $ 2.4 \times 10^{-2}$ & $ 1.1 \times 10^{-3}$ & $ 1.3 \times 10^{-3}$ & $ 6.2 \times 10^{-4}$ \\
$log(S)<-15.5$ & $ 1.9               $ & $ 1.7               $ & $ 2.4               $ & $ 0.2               $ & $ 0.2               $ & $ 0.3               $ & $ 1.8 \times 10^{-2}$ & $ 2.2 \times 10^{-2}$ & $ 1.1 \times 10^{-2}$ \\
$log(S)<-16.0$ & $ 7.6               $ & $ 4.8               $ & $21.8               $ & $ 0.6               $ & $ 0.4               $ & $ 2.8               $ & $ 0.1               $ & $ 0.1               $ & $ 0.2               $ \\
$log(S)<-16.5$ & $21.9               $ & $10.7               $ & $ 1.6 \times 10^{2} $ & $ 1.9               $ & $ 1.0               $ & $20.1               $ & $ 0.5               $ & $ 0.5               $ & $ 1.5               $ \\
$log(S)<-17.0$ & $63.0               $ & $23.0               $ & $ 9.6 \times 10^{2} $ & $ 5.7               $ & $ 2.2               $ & $ 1.2 \times 10^{2} $ & $ 1.6               $ & $ 1.1               $ & $10.5               $ \\
$log(S)<-17.5$ & $ 1.7 \times 10^{2} $ & $46.3               $ & $ 5.3 \times 10^{3} $ & $15.5               $ & $ 4.5               $ & $ 6.7 \times 10^{2} $ & $ 4.6               $ & $ 2.5               $ & $65.1               $ \\
$log(S)<-18.0$ & $ 4.6 \times 10^{2} $ & $92.6               $ & $ 2.8 \times 10^{4} $ & $39.5               $ & $ 8.8               $ & $ 3.6 \times 10^{3} $ & $12.3               $ & $ 5.2               $ & $ 3.5 \times 10^{2} $ \\
$log(S)<-18.5$ & $ 1.2 \times 10^{3} $ & $ 1.8 \times 10^{2} $ & $ 1.4 \times 10^{5} $ & $93.3               $ & $16.3               $ & $ 1.7 \times 10^{4} $ & $31.6               $ & $10.1               $ & $ 1.7 \times 10^{3} $ \\
\hline
\multicolumn{10}{c}{$7.5<z<8.5$} \\
\hline
$log(S)<-14.0$ & $ 3.5 \times 10^{-4}$ & $ 3.2 \times 10^{-4}$ & $ 1.2 \times 10^{-4}$ & ---     & ---     & ---     & ---     & ---     & ---     \\
$log(S)<-14.5$ & $ 1.0 \times 10^{-2}$ & $ 1.0 \times 10^{-2}$ & $ 3.7 \times 10^{-3}$ & $ 2.6 \times 10^{-4}$ & $ 3.2 \times 10^{-4}$ & $ 1.6 \times 10^{-4}$ & ---     & ---     & ---     \\
$log(S)<-15.0$ & $ 0.2               $ & $ 0.2               $ & $ 7.6 \times 10^{-2}$ & $ 5.1 \times 10^{-3}$ & $ 6.2 \times 10^{-3}$ & $ 3.3 \times 10^{-3}$ & ---     & ---     & ---     \\
$log(S)<-15.5$ & $ 1.3               $ & $ 1.2               $ & $ 1.1               $ & $ 2.9 \times 10^{-2}$ & $ 3.1 \times 10^{-2}$ & $ 4.5 \times 10^{-2}$ & ---     & ---     & ---     \\
$log(S)<-16.0$ & $ 4.4               $ & $ 3.2               $ & $10.6               $ & $ 0.1               $ & $ 9.6 \times 10^{-2}$ & $ 0.4               $ & $ 7.5 \times 10^{-4}$ & $ 8.7 \times 10^{-4}$ & $ 4.3 \times 10^{-4}$ \\
$log(S)<-16.5$ & $15.2               $ & $ 7.9               $ & $81.0               $ & $ 0.3               $ & $ 0.2               $ & $ 3.4               $ & $ 1.3 \times 10^{-2}$ & $ 1.6 \times 10^{-2}$ & $ 7.6 \times 10^{-3}$ \\
$log(S)<-17.0$ & $44.2               $ & $17.2               $ & $ 5.2 \times 10^{2} $ & $ 1.0               $ & $ 0.5               $ & $21.5               $ & $ 9.7 \times 10^{-2}$ & $ 0.1               $ & $ 9.9 \times 10^{-2}$ \\
$log(S)<-17.5$ & $ 1.2 \times 10^{2} $ & $34.7               $ & $ 2.9 \times 10^{3} $ & $ 2.5               $ & $ 0.9               $ & $ 1.2 \times 10^{2} $ & $ 0.4               $ & $ 0.4               $ & $ 0.9               $ \\
$log(S)<-18.0$ & $ 3.1 \times 10^{2} $ & $68.6               $ & $ 1.6 \times 10^{4} $ & $ 7.2               $ & $ 1.8               $ & $ 6.6 \times 10^{2} $ & $ 1.2               $ & $ 0.9               $ & $ 6.4               $ \\
$log(S)<-18.5$ & $ 7.7 \times 10^{2} $ & $ 1.3 \times 10^{2} $ & $ 7.9 \times 10^{4} $ & $17.5               $ & $ 3.5               $ & $ 3.3 \times 10^{3} $ & $ 3.4               $ & $ 2.0               $ & $36.6               $ \\
\hline
\multicolumn{10}{c}{$8.5<z<9.5$} \\
\hline
$log(S)<-14.0$ & $ 1.2 \times 10^{-4}$ & $ 1.1 \times 10^{-4}$ & ---     & ---     & ---     & ---     & ---     & ---     & ---     \\
$log(S)<-14.5$ & $ 4.2 \times 10^{-3}$ & $ 4.1 \times 10^{-3}$ & $ 1.5 \times 10^{-3}$ & ---     & ---     & ---     & ---     & ---     & ---     \\
$log(S)<-15.0$ & $ 9.9 \times 10^{-2}$ & $ 0.1               $ & $ 3.1 \times 10^{-2}$ & $ 6.0 \times 10^{-4}$ & $ 8.2 \times 10^{-4}$ & $ 4.3 \times 10^{-4}$ & ---     & ---     & ---     \\
$log(S)<-15.5$ & $ 0.7               $ & $ 0.7               $ & $ 0.5               $ & $ 4.6 \times 10^{-3}$ & $ 5.6 \times 10^{-3}$ & $ 7.0 \times 10^{-3}$ & ---     & ---     & ---     \\
$log(S)<-16.0$ & $ 3.1               $ & $ 2.3               $ & $ 5.5               $ & $ 2.1 \times 10^{-2}$ & $ 1.9 \times 10^{-2}$ & $ 7.7 \times 10^{-2}$ & ---     & ---     & ---     \\
$log(S)<-16.5$ & $10.1               $ & $ 5.7               $ & $44.9               $ & $ 6.4 \times 10^{-2}$ & $ 4.5 \times 10^{-2}$ & $ 0.6               $ & ---     & ---     & ---     \\
$log(S)<-17.0$ & $28.4               $ & $12.4               $ & $ 2.9 \times 10^{2} $ & $ 0.2               $ & $ 0.1               $ & $ 4.0               $ & $ 2.2 \times 10^{-3}$ & $ 2.6 \times 10^{-3}$ & $ 1.2 \times 10^{-3}$ \\
$log(S)<-17.5$ & $77.7               $ & $25.3               $ & $ 1.7 \times 10^{3} $ & $ 0.6               $ & $ 0.2               $ & $23.5               $ & $ 2.7 \times 10^{-2}$ & $ 3.3 \times 10^{-2}$ & $ 1.7 \times 10^{-2}$ \\
$log(S)<-18.0$ & $ 2.3 \times 10^{2} $ & $53.6               $ & $ 9.3 \times 10^{3} $ & $ 1.3               $ & $ 0.4               $ & $ 1.3 \times 10^{2} $ & $ 0.2               $ & $ 0.2               $ & $ 0.2               $ \\
$log(S)<-18.5$ & $ 5.6 \times 10^{2} $ & $ 1.0 \times 10^{2} $ & $ 4.8 \times 10^{4} $ & $ 4.2               $ & $ 0.9               $ & $ 6.5 \times 10^{2} $ & $ 0.5               $ & $ 0.5               $ & $ 1.5               $ \\
\hline
\end{tabular}
\caption{Predicted QSOs counts at different X-ray flux limits (deg$^{-2}$)}\label{tab:counts4}
\end{center}
\end{table*}

\end{document}